\documentstyle[preprint,tighten,aps]{revtex}
\preprint{KUNS 1345}
\draft
\title{Delta degrees of freedom in antisymmetrized molecular dynamics\\
and \\
(p,p$'$) reactions in the delta region}
\author{Andreas Engel, Eiji I. Tanaka, Tomoyuki
Maruyama\thanks{Present address: Advanced Science Research
  Center, Japan Atomic Energy Research Institute, Tokai-mura,
  Ibaraki-ken 319-11, Japan.},
Akira Ono\thanks{Present address: Cyclotron Laboratory, Institute of
Physical and Chemical Research (RIKEN), Hirosawa, Wako-shi,
Saitama-ken 351-01, Japan.}, and Hisashi Horiuchi}
\address{Department of Physics, Kyoto University, Kyoto 606-01,
Japan}
\begin{document}
\maketitle
\begin{abstract}
Delta  degrees of freedom are introduced into
antisymmetrized molecular dynamics (AMD). This is done by
increasing the number of basic states in the AMD wave function,
introducing a Skyrme-type
delta-nucleon potential, and including $NN\leftrightarrow N\Delta$
reactions in the collision description. As a  test of the
delta dynamics,  the
extended AMD  is applied to (p,p$'$)  recations
at $E_{\rm lab}=800$ MeV
for a $^{12}$C target.
It is found that the
ratio and the absolute values for delta peak and quasielastic peak
(QEP) in the $^{12}$C(p,p$'$) reaction are   reproduced for angles
$\Theta_{\rm lab} \agt 40^\circ$, pointing to a correct
treatment of the delta dynamics in the extended AMD. For forward
angles the QEP is overestimated, but in generally the agreement between
AMD calculations and experimental data
is good.  The results of the AMD calculations are compared
to one-step Monte Carlo (OSMC) calculations  and a detailed analysis of
multi-step
and delta potential effects is given. Along this analysis a
decompostion of the cross
section into various reaction
channels is presented and   the  reaction dynamics is discussed in
detail.
As important side results we present a way to apply a Gallilei invariant
theory for (N,N$'$) reactions up to $E_{\rm lab} \approx 800$ MeV which
ensures approximate Lorentz invariance and we discuss how to fix the
width parameter $\nu$  of  the single particle
momentum distribution  for
outgoing nucleons in the AMD calculation.
\end{abstract}
\pacs{25.40.Ep, 24.10.Cn, 14.20.Gk}

\narrowtext
\section{Introduction}
In the interesting and
 expanding field of heavy ion physics several models for the treatment
of the reaction dynamics have
been developed. Especially
microscopic transport models which take  nuclear mean
field effects and two particle collisions  into account have been very
successful in explaining various kind of experimental
data. These  models are extentions of the basic intra nuclear
model (INC) \cite{cug81,cug82} in
 which  two nucleon collisions are taken into account, but
the nuclear mean field is not treated self consistently.

Two major microscopic
approaches are  used:  transport
models of the Boltzmann-Uehling-Uhlenbeck type (BUU or VUU) for the nuclear
phase-space distribution function, solving either
a non relativistic transport equation
(non relativistic BUU) \cite{ber88,cas90,mos91}
or a relativistic transport equation (RBUU) \cite{bla88} and
 the quantum molecular dynamics (QMD) \cite{QMD} or
its relativistic version (RQMD) \cite{sor89}, in which
the nuclear
phase-space distribution function is modeled by the
sum of single-nucleon Gaussian wave packets, whose peak
positions are propagated in time according to the many-body
Newton equation, taking into account the self consistent
mean field  as the sum of all two-body potentials.

In this paper we will discuss  especially the antisymmetrized
molecular dynamics (AMD) \cite{ONO}  which is an extension  of the QMD
and is similar to the fermionic molecular dynamics (FMD) proposed by
Feldmeier \cite{FELD}.  In the AMD
the nuclear wave function is assumed to be
a Slater determinant of single-nucleon
Gaussian wave packets. Other than in the FMD the Gaussian width
parameter is  time independent and, in addition to the pure
mean field dynamics which is treated in the FMD, the two nucleon
collision process has been incorporated in the AMD \cite{ONO}.

The AMD   has   proved to be very successful in
describing heavy ion collisions at medium energies ( $\alt$200
MeV/nucleon)
\cite{ONO,ONOA,ONOB,ONOC,BODR,TAOR,VARR}.  Because the
 AMD describes the total
system with a Slater determinant of nucleon wave packets it
has quantum mechanical character, which has been demonstrated in the
ability of treating shell effects in the dynamical formation of
fragments.  Furthermore it has been shown that ground state wave
functions of colliding nuclei given by the AMD  are realistic and
reproduce many spectroscopic data very well
\cite{HORI,HORIA,NEKANADA,SANTO}.

Since the AMD is  successful for energies below 200 MeV/nucleon we want to
apply it also at higher energies Besides the question of Lorentz
invariance, which we will also comment on in this paper, the opening of
inelastic channels, i.e.\ the excitation of delta degrees of freedom,
 in the elementary nucleon-nucleon collision process
is the most important effect for $E_{\rm lab}/A \agt 600$ MeV. Therefore
the  first
part of this paper is  concerned with the question how to
incorporate these inelastic channels into the AMD framework.

Before applying  the extended AMD to heavy ion reactions we want to
get a good understanding of the basic dynamics in our model. We choose
to investigate  (p,p$'$)  reactions at delta resonance
energies. This is in the spirit of thr research done by Engel
et al.\ \cite{eng94}
who studied the pion-nucleus reaction to get a better  understanding
of  the pion and delta dynamics in
the BUU model.
Another reason to apply the extended AMD to (p,p$'$) reactions is that
we found in a previous paper \cite{enj95} that the AMD can
reproduce data for (p,p$'$) reactions for energies up to 200 MeV very
well, so we want to see if this is also true for higher energies.

It is to be noted that  the (p,p$'$) reactions  have been a major field of
interest \cite{ost92}, since other than in the (p,n) reaction the delta
peak is not shifted. Therefore we will investigate the delta potential
dependence and the reaction dynamics in detail.

Finally to get a better understanding of the rather elaborated AMD calculation
for the (p,p$'$)  reaction and the experimental data we compare to a
one-step Monte Carlo (OSMC)
simulation for  this process. Thereby we can study the influence and
importance  of mean field effects, multi-step processes, and Lorentz
invariant kinematics.

The organization of this paper is as follows: In Sec.\ \ref{amd} we explain
the AMD framework, discussing  the
introduction of delta degrees of freedom in the mean field  and
in the collision part; i.e.\ the  delta potential and $NN\rightarrow
N\Delta$ cross sections
are introduced. In this section we also discuss   the first
calculations of delta separation  energies of
'delta nuclei' in the AMD.
 In Sec.\ \ref{monte}  we  give a description of the OSMC which we use
for comparison with the AMD
calculations. As an important basis for the studies presented in this
paper, we study the question of  Lorentz invariance for the  (p,p$'$)
reaction within the OSMC. We find and show the interesting result,
that a  combination
of Lorentz transformation from the laboratory frame into the total
equal-velocity frame, Gallilei
invariant treatment in the equal-velocity frame, and following Lorentz
back transformation into the laboratory frame ensures approximate
Lorentz invariant results for energies around $E_{\rm lab} \approx
800$ MeV.
  In Sec.\ \ref{exper} we discuss  the results of AMD and  OSMC in
several steps and compare them with one another and with experimental
data. First we present
the results of the OSMC calculations and compare them with experimental
data. Before  we discuss the improvements which are achieved by
the full AMD calculation, we explain how to fix all the parameters,
especially the width parameter $\nu$ in the single particle momentum
distribution of the outgoing nucleons, in the AMD cross section
calculation independent of the experiment. This is an important point
of the calculation  and should be understood well.  After a detailed
discussion
of the multi-step contributions to the cross section in the AMD calculation
we proceed to discuss  multi-step and  potential effects by comparing
OSMC and AMD calculation with one another. In a final step we discuss
dependence on delta
potential, on  elastic nucleon-nucleon cross section and
on target momentum distribution of the calculations,  and possible
improvements of the AMD
 calculations.
In Sec.\ \ref{summ} we give the summary and an outlook on what should or
what could be done in future.

\section{Formulation of extended AMD}
\label{amd}
\subsection{Basic idea}
\label{amdidea}
For intermediate energy collisions  E$_{\rm lab}$/A $\agt$ 600 MeV
inelastic channels open in
the elementary nucleon-nucleon process. The first channel to open is
the pion production. As it is kown from one-boson exchange model
calculations for the elementary process $NN\rightarrow NN\pi$
\cite{ver79,sch93}
 the pion
production proceeds mainly through excitations of delta states
as it is depicted  in the Feynman diagram in
Fig.\ \ref{Fnnnnpi}.
 Diagrams
with an excitation of an off-shell nucleon are only important for
energies near the pion production threshold.

To include additional processes in the collision part of microscopic
models the cross sections for this processes are needed. In the pion
production case these cross sections are well known from experiment,
but still
there are in principle two ways to include pion production  into
microscopic models.  One way is to treat the $NN\rightarrow NN\pi$
reaction directly, but this approach has several deficiencies. First
of all the pions in heavy ion collisions are produced in the nuclear
medium and therefore can be reabsorbed during the collision. Since
 the pion absorption process in the nucleus is not well
understood, it is not clear how to treat it after a pion productuion
in the nuclear medium.
 Salcedo et al.\ \cite{sal88} calculated density-dependent pion absorption
probabilities in the nucleus, but these calculations are based on local
Thomas Fermi approximation and therefore are only valid for
ground state nuclei and cannot be used in heavy ion collisions. Another
drawback of the pure pion and nucleon cascade approach is that
it is well known from experiment  that also delta degrees of
freedom survive in the nuclear medium.

Due to this drawbacks we choose another way which is also adopted in
other microscopic models (see  for example Refs.\
\cite{cug81,sor89,mol85,wol90}).
We cut the $NN\rightarrow NN\pi$ reaction
into two steps: 1) $NN\rightarrow N\Delta$ and 2) $\Delta \rightarrow
N\pi$ reaction. This is possible since the decay of deltas into pions
occupies
nearly 100\% of the delta width.

To simulate the produced pion and nucleon momentum spectra correctly
 we
have to take  a mass distribution of the intermediate
excited delta  resonance into account. The origin of this mass distribution is
easily understood by calculating the contribution of the Feynman diagram in
Fig.\ \ref{Fnnnnpi}
to the pion production cross section. In the production cross section
a term proportional to the square of the
delta propagator
\begin{equation}
\label{deltaprop}
G_\Delta \sim  \frac{1}{s-(M_\Delta-{\rm i} \Gamma(s)/{2})^2}
\end{equation}
will appear, where $s$ is the square of the four momentum of the
delta, $M_\Delta$ the delta rest mass and
 $\Gamma (s)$ the width of the delta (see Eq.\  (\ref{gamma})).
 Therefore off-shell states will contribute to the cross section
with  the weight of  a $s$-dependent factor proportinal to $G_\Delta^2$. To
simulate the off-shell character of the delta in microscopic models
we populate delta states
with a mass distribution proportional to  $G_\Delta^2$, interpreting
$s$ as mass of  the delta, but propagating the delta states as
on-shell particles (see Sec.\ \ref{amdcoll}  ).

Now by including also the reverse processes $N\Delta\rightarrow NN$
and $\pi N\rightarrow \Delta$ we get a consistent picture of the delta
and pion dynamics in heavy ion collisions. Unklike  in the pure
pion case discussed above, the pion absorption process
is fixed in this approach,
namely, 1) $\pi N\rightarrow \Delta$ and 2) $N\Delta\rightarrow NN$
lead to a pion absorption. It has been shown that this treatment of
the collision term can reproduce many experimental data
for the   pion-nucleus reactions
quite well \cite{eng94}. Also unlike  in the pure pion case
the delta is included explicitly as degree of
freedom.

As  a first step of extending  the AMD we  discuss in this paper
only the extension to include  delta  degrees of freedom. We find that for
calculations of (p,p$'$) reactions for light targets  the treatment of
only delta degrees of freedom is sufficient. The decay of the deltas
is calculated in the final time step of the calculation, when the
deltas have escaped the the nuclear medium. Later we plan to include also
pion degrees of freedom in the nuclear medium and the full pion
dynamics as explained above.

Besides the inclusion of the delta in the collision part it also has
to be included into the mean field part of the theory. Normally this is
done by treating the delta as heavy nucleon and propagating it in the
same mean field as the one for  nucleons. More refined
 calculations have been
done which use different mean potentials for deltas and
nucleons \cite{ehe93}. We adopted also a delta potential which is
different from the nucleon potential.

We will proceed in the following Sec.\  \ref{amdmean}
by explaining how to describe a system
of deltas and nucleons in AMD, presenting  the basic formulas and also the
various potentials used in the later  calculations. After this we
describe in Sec.\ \ref{amdcoll} the adopted  cross sections, i.e.\ the
delta related cross sections, and how we
treat the collision term   in the
extended AMD.

\subsection{Mean field with delta}
\label{amdmean}

The details of the AMD formulation have been  explained in
Ref.\ \cite{ONO}. We will only comment on the modifications  due to
the introduction of delta degrees of freedom and those points which
are important for the understanding of the  calculation discussed in
Sec.\ \ref{exper}.

In AMD, the wave function of $A$-nucleon system is described by a
Slater determinant $|\Phi(Z)\rangle$,
\begin{equation}
|\Phi(Z)\rangle = {1\over\sqrt{A!}}\det
\Bigl[\varphi_j(k)\Bigr], \quad
\varphi_j = \phi_{{\bf Z}_j} \chi_{{\alpha}_j}
\label{AMDWF}
\end{equation}
where $\chi$ stands for the spin-isospin function and $\alpha_j$
represents the spin-isospin label of the $j$-th single particle state,
$\alpha_j={\rm p}\uparrow$, ${\rm p}\downarrow$, ${\rm n}\uparrow$, or
${\rm n}\downarrow$.  $\phi_{{\bf Z}_j}$ is the spatial wave function
of the $j$-th single-particle state which is a Gaussian wave packet,
\begin{equation}
\begin{array}{rcl}
 \langle {\bf r} | \phi_{{\bf Z}_j} \rangle & = &
\displaystyle
 \Bigl( {2\nu \over \pi} \Bigr)^{3/4} \exp \Bigl[ -\nu \Bigl(
 {\bf r} - {{\bf Z}_j \over {\sqrt \nu}} \Bigr)^2 +
      {1 \over 2}{\bf Z}^2_j \Bigr], \\
 {\bf Z}_j  & = & \displaystyle{\sqrt{\nu} {\bf D}_j +
   {i \over 2\hbar\sqrt{\nu}} {\bf K}_j},
\end{array}
\label{GAUSSWF}
\end{equation}
where the width parameter $\nu$ is treated as time-independent in the
present work.  We take $\nu$=0.16 fm$^{-2}$ in the calculation in this
paper.  Here ${\bf Z}_j$ is the complex vector whose real and
imaginary parts, ${\bf D}_j$ and ${\bf K}_j$, are the spatial and
momentum centers of the packet, respectively.

For the description of delta states we extend the possible number of
spin-isospin  functions $\chi$ in Eq.\ (\ref{AMDWF}). This leads to
the introduction of 16
additional states. But since we  only use spin-independent cross sections
and spin-independent delta
potential we can reduce the number of delta states to four,
\begin{equation}
\alpha_j^\Delta = \Delta^{++},\Delta^{+},\Delta^{0},\Delta^{-} \quad ,
\end{equation}
or in other words we use a delta spin averaged wave function.
For deltas we use the same spatial wave functions as for nucleons
as given in Eq.\ (\ref{GAUSSWF})
leaving also the width parameter $\nu$ unchanged.

In the same way as in the original AMD
the  time developments of the coordinate parameters, $Z = \lbrace {\bf
Z}_j \ ( j=1,2, \ldots, A )\rbrace$, due to mean field
propagation is determined by the time-dependent
variational principle;
\begin{eqnarray}
  \delta \int_{t_1}^{t_2} dt
  { \langle \Phi(Z) | \Bigl(
  i\hbar {\displaystyle {d\over dt}}-H \Bigr) |
  \Phi(Z) \rangle  \over \langle \Phi(Z) |
  \Phi(Z) \rangle } = 0,
\label{eq}
\end{eqnarray}
which leads to the equation of motion for $Z$,
\begin{equation}
\begin{array}{rcl}
&& \displaystyle{
i\hbar \sum_{j\tau} C_{k\sigma,j\tau} {d\over dt} Z_{j\tau} =
  {\partial \over \partial {Z_{k\sigma}^*}}
  { \langle \Phi(Z) | H | \Phi(Z) \rangle
  \over \langle \Phi(Z) | \Phi(Z) \rangle}}, \\[0.4cm]
&& \displaystyle{
C_{k\sigma,j\tau}
  \equiv {\partial^2\over\partial Z_{k\sigma}^*\partial Z_{j\tau}}
      \ln \langle \Phi(Z) | \Phi(Z) \rangle} ,
\end{array}
\label{EQMTN}
\end{equation}
where $\sigma, \tau=x,y,z$.

In the case of the extended AMD the many-body Hamiltonian has to be
changed. For a A-body system with $N_N$ nucleons and $N_\Delta$ deltas
($A=N_N +N_\Delta$) the Hamiltonian is given by
\begin{eqnarray}
\label{HAMIL}
H &=& T + V \nonumber\\[0.3cm]
&= & \sum_{i_N=1}^{N_N} \frac{{\bf p}_{i_N}^2}{2m} +
\sum_{i_\Delta=1}^{N_\Delta}
 \frac{{\bf p}_{i_\Delta}^2}{2m_{i_\Delta}}
+ \sum_{i_N < j_N}^{N_N} v_{NN}(i_N,j_N) \\
& &
+ \sum_{i_N}^{N_N} \sum_{i_\Delta}^{N_\Delta}
v_{N\Delta}(i_N,i_\Delta)
+ \sum_{i_\Delta < j_\Delta}^{N_\Delta}
v_{\Delta\Delta}(i_\Delta,j_\Delta)\quad .\nonumber
\end{eqnarray}
It is a sum of kinetic energies for deltas and nucleons and the
various two-body potentials for nucleon-nucleon, nucleon-delta and
delta-delta interaction.

During the dynamical reaction stage of the collision, the total system
can be separated
into several isolated nucleons and fragments.  Since the wave
functions of the center-of-mass motion of these isolated nucleons and
fragments are Gaussian wave packets, each of these isolated particles
carries spurious zero-point energy of its center-of-mass motion.  The
total amount of the spurious energy of center-of-mass motion can be
expressed as a function of $Z$ \cite{ONO,ONOB}, which we denote as
$E_{\rm sprs}(Z)$.  The actual Hamiltonian we use in the above
equation of motion (Eq.\ (\ref{EQMTN})) is therefore given by
$\langle \Phi(Z) | H | \Phi(Z) \rangle /\langle \Phi(Z) | \Phi(Z)
\rangle - E_{\rm sprs}(Z)$.

As discussed in Sec.\ \ref{amdidea} the deltas in the AMD will have
a different mass than the free on-shell mass. This is reflected in
Eq.\ (\ref{HAMIL}) by using $m_{i_\Delta}$ in the expression for the
kinetic energy.

Eq.\ (\ref{EQMTN}) with definitions Eqs.\ (\ref{HAMIL}) and
(\ref{AMDWF})
are the
basic formula for the extended AMD on behalf of the mean field,
 but to do actual calculation the form of the potentials have to be
fixed.

For  the effective two-nucleon force we adopt the Gogny force
\cite{GOGNY} which has been successfully used in studying heavy ion
reactions with AMD \cite{ONOB,ONOC}.  The Gogny force consists of
finite-range two-body force and density-dependent zero-range repulsive
force. This force gives a momentum-dependent mean field which
reproduces well the observed energy dependence of the nucleon optical
potential up to about 200 MeV but levels off to be zero at higher
energies
whereas the
experimental value at $E_{\rm lab}=$ 800 MeV is repulsive by about 50 MeV.
The nuclear matter EOS given by Gogny
force is soft with the incompressibility $K = $ 228 MeV. Corresponding
to the choice of Gogny force, the calculational formula of the total
spurious center-of-mass energy $E_{\rm sprs}(Z)$ is taken to be the
same as Ref.\ \cite{ONOB}.  The binding energies of $^{12}$C is
 calculated to be 92.6 MeV while the
observed value is 92.2 MeV.  The
calculated r.m.s.\ radii of $^{12}$C  is 2.55 fm
which is reasonable.

For the delta-nucleon potential we use a zero-range
Skyrme-type  force dependent on  nucleon
density ($\rho_N$),
\begin{equation}
v_{N\Delta}({\bf r}_N,{\bf r}_\Delta  ) = \delta
({\bf r}_N -{\bf r}_\Delta ) [ t_{0\,
\Delta} +
t_{3\, \Delta}
\rho_N^{\tau -1}({\bf r}_N)]
\label{DELTPOT}
\end{equation}
with parameters $t_{0\, \Delta}, t_{3\, \Delta}$ and $\tau$.
 This  ansatz leads to a total
delta-nucleus energy
\begin{eqnarray}
V_\Delta &=&   \langle \Phi(Z) |  \sum_{i_N}^{N_N} \sum_{i_\Delta}^{N_\Delta}
v_{N\Delta}(i_N,i_\Delta)
 | \Phi(Z) \rangle /
\langle \Phi(Z) |   \Phi(Z) \rangle \\
\label{DELTTOT}
&= &
\int d{\bf r} \,\,
\rho_\Delta({\bf r}) \,\, \Bigl(t_{0\, \Delta} \rho_N({\bf r}) +
   t_{3\, \Delta} \rho_N({\bf r})^\tau  \Bigr) \quad ,
\end{eqnarray}
where $\rho_N$ and $\rho_\Delta$ are the nucleon  density and delta
density, respectively. If we take the limit of infinite homgenous
nuclear matter of Eq.\ (\ref{DELTTOT}), we get the same form of  the
delta potential derived from the delta hole model by Ehehalt
 et al.\ in Ref.\ \cite{ehe93}.
Therefore we use their parameters, $t_{0\, \Delta}$= $-700$ MeV fm$^3$, $t_{3\,
\Delta}$=
1750 MeV fm$^5$ and  $\tau = 5/3$ in the calculation. The parameters
are  chosen to reproduce the mean field for deltas $U_\Delta (\rho_0)=
\partial V_\Delta (\rho_0 )
/ \partial \rho_\Delta  =
 -30$ MeV for infinte nuclear matter, since this value is known from
pion-nucleus scattering \cite{hor80}.

It is important to note that the nucleon-nucleon potential we use has
finite-range which leads to a
momentum-dependent nucleon-nucleus potential. On the other side  the
delta-nucleon
potential adopted here  has zero-range,
and therefore the delta-nucleus potential is
momentum-independent.

Since there is nothing known about the delta-delta potential we
  choose $v_{\Delta ,\Delta}$ to be zero. As a remark we want to point
out that the calculation presented in this paper are independent of
the choice of the delta-delta potential because in the calculated
nucleon-nucleus
reactions
 there is a maximum of one delta excited at a time. For the
future calculation of heavy ion collisions we propose to use a
delta-delta potential of the same type as the delta-nucleon potential
Eq.\ (\ref{DELTPOT}) since these potentials should be similar
 except for spin-isopin factors  and
this choice gives a consistent picture
of the interaction.

Numerically the integrations in Eq.\ (\ref{DELTTOT}) to calculate
$V_\Delta (|\Phi({\bf Z})>)$ and those needed to get
its  derivatives due to ${\bf Z}_i$ are calculated in a similar  way as
those described in the Appendix A of \cite{ONOA} for the
density-dependent
 part of the nucleon-nucleon interaction.

We also include the Coulomb potential in our calculation, but the
results are not sensitive to this.

\subsection{Delta nuclei}
\label{amdnuc}
To construct the ground states of colliding nuclei in the AMD the
frictional cooling method \cite{HORI,HORIA,NEKANADA,SANTO} is used
which leads
to a consistent picture of mean field dynamics and ground state nuclei. It has
been checked that wave functions given by AMD are realistic and
reproduce many spectroscopic data very well.

As the first check of the incorporated delta mean field dynamics we
calculated the binding energies of 'delta nuclei'. This is in a way an academic
study since no 'delta nuclei' exist or can be detected, but it gives
us information on the
kind of wave functions we can have in a delta-nucleus system. In future
this kind of wave functions could be used for
studying the admixture of delta states in ground state wave function
\cite{ame94}.

We calculated the delta separation  energies for 'delta nuclei'.
For this
we used the frictional cooling method to cool first  an
ensemble of protons, neutrons and one additional delta.
Then   we cooled
the same ensemble of nucleons without delta.
 The difference of the minimun
energies of these two calculations is  the given delta separation
energy. The results for delta
separation
energies with the parameters given  in Sec.\ \ref{amdmean}
are shown  in Fig.\ \ref{Fdeltbind}.

The results in Fig.\ \ref{Fdeltbind} can be understood in the
following way:
The delta separation energy is determined by the overlap between the
delta Gaussian wave function and the residual many-nucleon system, as
can be seen in Eq.\ (\ref{DELTTOT}). Therefore the minimum energy state
calculated with the frictional cooling method leads to a delta state with
maximum overlap with the nucleons. Because of the 3/2 isospin nature of the
delta the configuration of maximum overlap
is Pauli allowed. The mass dependence  we get
 in Fig.\
\ref{Fdeltbind} for the delta separation energy reflects the structure
of the  ground state wave function in  AMD, which was able to describe
the alpha clustering of light nuclei. We found that the delta
separation energies for $^4$He and
 $^8$Be are the same.  This can be understood on the basis of the two
alpha clustering of the  $^8$Be ground state. In the ground states of
both nuclei the delta is in the middle
of an alpha cluster, since this is the state of maximum overlap of
delta and nucleus wave functions. Therefore they have the same delta
separation  energy. Because of the disolving of the alpha
structure and the existence of larger zones with higher
nuclear densities for heavier masses, also the delta separation energies
increases for heavier masses.

\subsection{Deltas in the collision term}
\label{amdcoll}
The idea how to  incorporate  delta resonances  in the collision part
of microscopic theories was already
discussed in Sec.\ \ref{amdidea}. Here we give the cross sections
adopted in the AMD. To include the deltas in the
collision term we  used the same cross
sections described in \cite{eng94,wol90}, therefore
we will only recall the most
important formulas.

\subsubsection{$NN \rightarrow N\Delta$ reaction}
\label{NNND}
The cross sections for the $NN\rightarrow N\Delta$ reaction are
derived from the VerWest-Arndt parametrisation for pion
production cross sections  in Ref.\ \cite{ver82}. Asuming that
all pions in the $NN\rightarrow NN\pi$ reaction are produced via a
delta resonance we get
\begin{equation}
\begin{array}{lcl}
p + p \rightarrow  n + \Delta ^{++}&& \sigma _{10} +1/2
\sigma _{11}\\
p + p \rightarrow  p + \Delta ^+&&3/2 \sigma _{11}\\
n + p \rightarrow  p + \Delta ^0&&1/2 \sigma _{11} + 1/4
\sigma _{10}\\
n + p \rightarrow  n + \Delta ^+&&1/2 \sigma _{11} + 1/4
\sigma _{10}\\
n + n\rightarrow  p + \Delta ^-&&\sigma _{10} +1/2
\sigma _{11}\\
n + n \rightarrow  n + \Delta ^0&& 3/2\ \sigma _{11} \quad ,
\end{array}
\end{equation}
for the cross sections where $\sigma_{if}$
denotes the cross sections used  in the VerWest-Arndt
parametrisation  for given isospin  $(f)$ and $(i)$ of  the final and
initial nucleons, respectively \cite{ver82}.
The angular dependence of the $NN\rightarrow N\Delta$ is  chosen in the
same way as in Ref.\ \cite{wol90}.

As discussed in Sec.\ \ref{amdidea} it is not enough to define the
cross section for the $NN\rightarrow N\Delta$ reaction but we also
need to use an
appropriate mass distribution for the delta resonance. We choose
\begin{eqnarray}
\frac{ d\sigma_{NN\rightarrow N\Delta}}{dM^2} &=&
\sigma_{NN\rightarrow N\Delta}
\frac{F(M^2)}{\int_{(m_N+m_\pi)^2}^{(\sqrt{s}-m_N)^2}F(M^2)dM^2} \quad
, \\
\label{massdist}
F(M^2) &=& \frac{1}{\pi} \frac{M_\Delta \Gamma (M)}
{(M^2-M_\Delta^2)^2 + M_\Delta^2\Gamma (M)^2}\\
\label{massdist2}
  & \approx &
 \frac{1}{\pi M_\Delta} \frac{\Gamma (M)/4}
{(M-M_\Delta)^2 + \Gamma (M)^2/4}\nonumber\\
\end{eqnarray}
with the delta rest mass $M_\Delta$=1232 MeV and the momentum-dependent
delta width \cite{koc84}
\begin{equation}
\Gamma (M) =  \left(\frac{q}{q_r}\right)^3 \frac{M_\Delta}{M} \left(
\frac{v(q)}{v(q_r)}\right)^2\Gamma_r
\label{gamma}
\end{equation}
with
\begin{eqnarray}
 v(q)& = &\frac{\beta^2}{\beta^2 + q^2}\quad ,\nonumber\\
\Gamma_r &=& 110\, {\rm MeV} \quad ,\nonumber\\
 \beta &= &300\, {\rm
MeV} \quad ,\nonumber
\end{eqnarray}
where   $q$ and $q_r$ denote the pion momentum in the rest frame of
the delta for  delta mass $M$ and on-shell mass $M=M_\Delta$,
respectively, defined as
\begin{equation}
q^2(M) = \frac{[M^2-(m_N-m_\pi)^2][M^2-(m_N+m_\pi)^2]}{4M^2}
\quad .
\end{equation}

To get information on the validity  of the adopted delta mass distribution
we calculated the neutron momentum spectrum in the p(p,n)p$\pi^+$ reaction
and compared
to  the experimental shape of the spectrum. In
Fig.\ \ref{FpHn} the result of this calculation is shown. The
experimental data is a measurement of the neutron spectrum for laboratory
angles $\Theta_n = 0^\circ - 6^\circ$
in
coincidence with a proton and a $\pi^+$ for proton energy
$E_{\rm lab}=830$ MeV \cite{chi91}. This reaction proceeds mainly via an
excitation of a $\Delta^{++}$ state.
The calculations which are shown were made with OSMC
explained in detail in Sec.\ \ref{monte} using a proton target: A
delta excitation in a proton-proton  collision is simulated
choosing the mass and momentum according to the given
mass distribution (Eq.\ (\ref{massdist2})) and the scattering angle according
to the differential cross section of Ref.\ \cite{wol90}, respectively, by Monte
Carlo method.  The deltas are  assumed to decay isotropically in
their rest frame. Taking into account the proper transformations of
the momenta we
calculate the cross section for neutrons emitted in coincidence with a
$\pi^+$ and a proton.

Besides the calculation with the mass distribution of Eq.\ (\ref{massdist2})
we made calculations with a mass distribution multiplied by a
phase space factor
\begin{equation}
\label{factor}
f_{\rm phasespace}(M)=(\frac{p_\Delta(M)}{p_\Delta(M)^{max}})^n
\end{equation}
with
\begin{equation}
p^2_\Delta(M) = \frac{[s-(m_N-M)^2][s-(m_N+M)^2]}{4s} \quad ,
\end{equation}
where $s$ is the square of the four momentum of the system.
$p_\Delta(M)$ is the momentum of the delta and  $p_\Delta(M)^{max}$
is the possible maximum
momentum of the delta assumed for $M=m_n+m_\pi$.  With this phase
factor (Eq.\ (\ref{factor})) we take
into account the fact that the
 final phase space for maximun delta mass, $M=\sqrt{s}-m_N$,  is zero.
In  Fig.\ \ref{FpHn} calculations for the p(p,n)p$\pi^+$ reaction
with $n=0$, $n=2$ and $n=3$ in Eq.\ (\ref{factor}) are
shown. All calculations are scaled with the same factor to reproduce
the experimental data.
By taking into account only the mass distribution without
phase space suppression factor, $n=0$, solid line, we find that the
neutron spectrum is not reproduced so well. Too many heavy
$\Delta^{++}$ with, consequently, low momentum neutron partners  are
formed in proton-proton collisions. Therefore the neutron momentum
spectrum is shifted to lower energies. On the other hand using a
strong suppression factor, $n=3$, leads to an underestimation
of low momentum neutrons as seen in Fig.\ \ref{FpHn} (dashed
line). The best fit for the neutron spectrum in the p(p,n)p$\pi^+$
reaction with the OSMC  simulation is attained when
using $n=2$ in
Eq.\ (\ref{factor}) for the suppression factor as seen in
Fig.\ \ref{FpHn} (dashed dotted line). Both low and high neutron momenta can
be described rather well. In Sec.\ \ref{monteexp} we will comment on the
dependence of the (p,p$'$) reaction on the choice of the mass distribution.

\subsubsection{$N\Delta \rightarrow NN$ reaction}
\label{NDNN}
Since there is no delta beam or delta target available the cross
section for the $\Delta N\rightarrow NN$ process has to be derived by
theory. It has been  pointed out by several authors
\cite{dan91,wol92,eng94} that the naive detailed balance formula is
not appropriate for this process. The simple argument to understand
this point is that in the nucleon-nucleon collisions in which the
deltas are formed the cross section has to be folded with the mass
distribution to get the mass of the delta in the simulation, as described
in Sec.\ \ref{NNND}, but in the $N\Delta \rightarrow NN$ reaction
the delta in the simulation has  a definite mass  and therefore the
mass distribution is not present for the calculation of the
$N\Delta \rightarrow NN$ cross
section. For a more detailed discussion of this problem see
Ref.\ \cite{eng94}.

For simplicity we adopted the formula derived by Wolf et
al.\ \cite{wol92}, which was proven to give a realistic description of
the $N\Delta\rightarrow NN$ cross section (see discussion in
Ref.\ \cite{eng94}). Therefore we use
\begin{equation}
\label{gyu}
 \sigma_{n\Delta^{++}\rightarrow pp}= \frac{1}{4}
\frac{p_N^2}{p_\Delta^2}
 \sigma_{pp\rightarrow n\Delta^{++}}
\frac{\int_{(m_N+m_\pi)^2}^{\infty} F(M^2)dM^2}
{\int_{(m_N+m_\pi)^2}^{(\sqrt{s}-m_N)^2} F(M^2)dM^2}\,\,
\end{equation}
with  the $\sigma_{NN\rightarrow N\Delta}$ cross sections from VerWest-Arndt
and the mass distribution $F(M^2)$ given in Eq.\ (\ref{massdist}).
 In Eq.\ (\ref{gyu})
 $p_\Delta$ and $p_N$ are the initial delta and the final nucleon
momenta  in the
center of mass frame, respectively,  whereas the factor
$1/4$ is due to  spin averaging and a symmetry factor
for identical particles in
the final state. Since this cross section (Eq. \ref{gyu}) is infinte
for zero energy collisons  we apply a low momentum cut off of
$\sigma^{max}=100$ mb, which  takes  the screening of the delta in the
nuclear medium into account. In case of the (p,p$'$) reaction the
screening is not
effective because the produced deltas are fast deltas.

\subsubsection{$NN$ elastic scattering}
For the nucleon-nucleon elastic scattering we used a new
parametrisation of Cugnon which is isospin dependent and therefore takes
into account proton-neutron  and proton-proton collisions
differently \cite{cug88}.
This is not the case
in the old isospin-independent Cugnon
parametrisation \cite{cug81}, which is mainly adopted  in other microscopic
models,  in which  all isospin channels are treated the same way,
in contrast to experiment. The isospin-dependent parametrisation also
includes a back-ward peak for the proton-neutron scattering which is
not included in the isospin-independent parametrisation.

\subsection{Simulation}
\label{sim}
The actual simulation of the AMD  is done in several steps. In the
first step the ground states of the target and projectile nuclei are
calculated by the frictional
cooling method. With these ground state configurations the collision
is calculated  in the CM frame (see discussion about Lorentz invariant
scattering in Sec.\ \ref{montelo}) using the above described cross
sections and mean fields for the particles. The delta decay is
not yet incorporated in the AMD therefore deltas will survive untill
the final state
of the calculation. The decay of these deltas is calculated asuming an
isotropic decay in the  rest frame of the delta.
 In this step the relativistic
expression for the pion energy is used.   In the  final step
 the statistical decay of primordial formed fragments is calculated.
  Primordial
fragments mean the fragments which are present when the dynamical
stage of the reaction has finished.  These fragments are not in their
ground states but are excited, and they decay through evaporation with
a long time scale.  In this paper, the switching time from the
dynamical stage to the evaporation stage was chosen to be 150 fm/c.
Statistical decays of fragments were calculated with the code of Ref.\
\cite{MARU} which is similar to the code of P\"uhlhofer \cite{PUHL}.

\subsection{Calculation of cross section}
The experimental data we want to compare with in this paper are given in the
laboratory frame. As discussed in Sec.\ \ref{montelo} we choose a
combination of Lorentz and Gallilei transformation to achieve
approximate Lorentz invariance  using a Gallilei invariant theory.
The combinations of transformations has to be also taken into account
for the cross section calculation.
In the equal-velocity frame the Gallilei invariant expression of the
double differential
cross section has the following form:
\begin{eqnarray}
&& {d^2\sigma \over d\Omega dp} =
  \int_0^\infty 2 \pi b db
  {d^2 {\cal N}( {\bf p}, b) \over d\Omega dp}   , \nonumber\\
&& {d^2 {\cal N}( {\bf p}, b) \over d\Omega dp} d\Omega dp
  = \biggl< \sum_{\scriptstyle i =\rm isolated\atop
                             \hfill\scriptstyle\rm protons}
 \rho_i( {\bf p}) \biggr>_b d^3p ,
\label{ANGDIS}\\
&& \rho_i( {\bf p})= |\langle {\bf p} | \phi_{{\bf Z}_i} \rangle|^2
  = \Bigl( {1 \over 2 \pi \hbar^2 \nu} \Bigr)^{3/2}
  \exp \Bigl[ - {1 \over 2 \hbar^2 \nu} ( {\bf p} - {\bf K}_i )^2
  \Bigr] \label{momentumdens}
\end{eqnarray}
where $d^3p=p^2dpd\Omega$ and $\langle\;\;\rangle_b$
stands for the average value over the events with impact
parameter $b$. In this formula, the outcoming
protons are expressed by Gaussian wave packets with momentum width
$\hbar \sqrt \nu$.

Applying the Lorentz transformation into the laboratory frame we get
\begin{equation}
{d^2\sigma \over d\Omega' dp'} =
  \int_0^\infty 2 \pi b db\biggl< \sum_{\scriptstyle i =\rm isolated\atop
                             \hfill\scriptstyle\rm protons}
 \rho'_i( {\bf p'}) \biggr>_b p'^2
\label{crosscal}
\end{equation}
with
\begin{equation}
 \rho'_i( {\bf p'}) =  \rho_i( {\bf p}) {E\over E'}
\end{equation}
where ${\bf p}$ and $E$ are the relativistic expressions  for the momentum
and the energy in the equal-velocity frame, ${\bf p'}$ and $E'$ the
relativistic momentum and energy
in the laboratory frame and $\rho_i( {\bf p})$ the function given in Eq.\
(\ref{momentumdens}).

\section{One-Step Monte Carlo  Simulation for (p,p$'$)}
\label{monte}

\subsection{Formulation}
\label{montefo}
In order to get a better understanding of the rather elaborate AMD
calculations, we also perform one-step Monte Carlo (OSMC) simulations of
the (p,p$'$)
reaction. In this section we explain the OSMC simulation. In the OSMC
simulation the summation over the impact parameter $\int 2 \pi b db$
is not made for simplicity.  The projectile protons are initialized in
the laboratory frame according
to the relativistic momentum energy relation. Since we studied the
influence of the target
 momentum distribution on the results for the (p,p$'$) reaction we
initialized  different target momentum distributions. As one
approach we attributed  a random momentum smaller than the Fermi momentum
at normal nuclear density
$|{\bf
p}|< |{\bf k_f}|$ to the nucleons. As another approach  we used the
local Thomas Fermi
ansatz to simulate the momentum distribution by first
choosing a position in coordinate space according to the space density
\begin{equation}
\rho(r) = \rho_0 \left[ 1+ \alpha \left(\frac{r}{a}\right)^2\right]
\exp \left[-\left(\frac{r}{a}\right)^2 \right]  \quad ,
\end{equation}
which is based on the harmonic-oscillator model, with parameters
$\alpha= 1.247$  and $a=1.649$ fm for the $^{12}$C target  determined
through experiment
\cite{jag74}. And   then we
attributed a momentum randomly smaller than the local
Fermi momentum, which gave us the momentum distribution $f(p)$,
\begin{equation}
f(p) = \int_{0}^{\infty}dr 4 \pi r^2 \rho(r)\frac{\Theta ( k_f(r) -
p)}{4\pi/3 (k_f(r))^3}\quad , k_f(r) = \left( \frac{3\pi^2}{2}\rho
(r)\right)^\frac{1}{3} \quad ,
\end{equation}
where $\Theta (x)$ is the step function; $\Theta (x)=1 $ for $x>0$,
 $\Theta (x)=0 $ for $x<0$.

As next step  we transform the energy and
the momentum of the projectile proton from the laboratory frame
into the equal-velocity frame of
target and projectile
with Lorentz transformation. The transformation $\beta$ for incident
proton in
z-direction  and a target with A nucleons with four momenta
\begin{eqnarray}
{\bf p_\mu} &=& \{E_p,0,0,p_z\}\\
{\bf p_\mu^A} &=& \{Am_n,0,0,0\}\\
\end{eqnarray}
into the equal-velocity frame with condition
\begin{equation}
\frac{p'_z}{E'_{p}} =- \frac{p'^A_z}{E'_A}
\end{equation}
is
\begin{equation}
\label{betaeq}
\{\beta_x, \beta_y,\beta_z\}=
\{0, 0, \frac{E_p}{p_z} -
\sqrt{ (\frac{E_p}{p_z})^2-1}\} \quad ,
\end{equation}
where A is the number of target nucleons and $m_n$ the nucleon mass.
It can be seen that this transformation is not dependent on the number
of target nucleons. This is natural since the total equal-velocity
frame is the same as the two nucleon CM frame, if we disregard the
Fermi motion of the target nucleons.

We choose to  tranform  into the total equal-velocity frame, because
in this frame non relativistic approximations can be applied.
In the equal-velocity frame we change to non relativistic kinematics using the
obtained momenta and the non relativistic expression for the
energy.  Also for the transformation of the target nucleon  momenta into
the  equal-velocity frame we use the non relativistic Gallilei
transformation.   In the next section it will be shown that this kinematical
treatment of the scattering, which we also adopt in the AMD, leads to
approximate Lorentz invariance.

The two nucleon collision in OSMC is calculated in the
CM frame of the two nucleons with
Gallilei invariant kinematics. As possible  collision processes we
incorporated
delta excitation and elastic scattering. We
fixed the ratio of delta excitation to elastic events beforehand,
which is well justified, since this ratio  is mainly determined by the
initial energy of the proton. The cross section ratios for $E_{\rm
lab}=800$ MeV which are incorporated  in the calculation are
given in Fig.\ \ref{Fcrossratio}.

In case of a delta event we have to determine the mass of the delta
and, since we take into account the charge of the particles, the
charge of the delta. The charge and mass are chosen according to the
ratio given
by the cross sections and the mass distribution in Sec.\ \ref{NNND},
respectively. To calculate the cross sections  the $\sqrt{s}$ of the two
nucleon system is needed.
This we calculate in the two nucleon Gallilei CM frame using the non
relativistic expression for the energies. In the next section the
validity of this evaluation is discussed.

The three momentum of the outgoing particles, deltas or nucleons, are
determined according to the
angle-dependent cross sections
 in the two nucleon CM by Monte Carlo method.
Since in the delta
excitation kinetic energy  is transformed into mass energy  the final
relative momentum in the CM frame for the
$NN\rightarrow N\Delta$ reaction decreases.

The decay of the excited deltas  is calculted in the
delta rest frame assuming an isotropic decay.   For the delta decay we
use a Lorentz invariant prescription because the pions are light
particles and need to be treated in relativistic invariant kinematics. It
is important to mention here that the Lorentz transformation in the
delta decay is only
adopted to transform from the delta rest frame to the equal-velocity
frame of the total system.
The charge of the pions and nucleons are chosen according the
probabilities given  by the
Clebsch Gordan coefficients for the decay.

To calculate the multiplicity distribution of the outgoing protons or
neutrons  in the  laboratory frame for given momenta and angle, the momenta
of the particles are back transformed from the total equal-velocity
frame to the laboratory frame by Lorentz transformation with ${
\beta_{\rm back}}$ = $-\beta$
given in Eq.\ (\ref{betaeq}).
The multiplicity distributions are calculated with the formula
\begin{equation}
\label{multdist}
\frac{d^2N}{dp_id\Omega_i} =  \frac{N_i}{\Delta p 2\pi \Delta
\cos\Theta N_{tp}}
\end{equation}
where  $N_i$ is  the number of nucleons with momentum $p_i-\Delta p/2\le p
\le p_i+\Delta p/2$  and $\cos\Theta_i-\Delta \cos\Theta/2\le \cos\Theta_i
\le\cos\Theta_i+\Delta \cos\Theta/2$, and $N_{tp}$  the number of
performed simulations.
  In the calculations shown we use  $\Delta p = 0.02$ GeV and
$\Delta \Theta = 2.99^\circ$. In  the calculation of cross sections for
the (p,p$'$) reaction we use   a factor  $X_{\rm
factor}$ to scale to the experimental
data:
\begin{equation}
\frac{d^2\sigma}{dp_id\Omega_i} =  \frac{d^2N}{dp_id\Omega_i}
X_{factor} \quad .
\end{equation}
The factor
 $X_{factor}$ has a dimension, which is because we do not integrate
 over the impact parameter.
 The number of simulations was varied according
to the accuracy needed.

\subsection{Lorentz invariance}
\label{montelo}
To justify the approach used for the scattering kinematics in the OSMC
simulations and also the AMD calculations we compared the OSMC simulation
described in
Sec.\ \ref{montefo} with simulations using pure Gallilei and pure Lorentz
invariant kinematics, respectively. For this comparison we did
one-step
elastic scattering simulations  using an isotropic angular distribution for the
scattering in the nucleon-nucleon CM frame disregarding the charges of
the particles. The results for the multiplicity distribution for
Lorentz invariant kinematics and OSMC simulations are shown in Fig.\
\ref{Florentz1}. The momentum of the target nucleons is chosen
randomly between zero and the Fermi momentum $k_f=270$ MeV.
 The calculations shown are for
$E_{\rm lab} = 800$ MeV and  scattering momenta in the laboratory
frame. We show calculations for laboratory
angles $\Theta = 15^\circ$ and $\Theta = 60^\circ$. The solid line represents
results with the
Lorentz invariant kinematics and the dashed line  results with the
OSMC.   The results agree very
well with each other. In Fig.\ \ref{Florentz2} the results of
Lorentz invariant calculations (solid line) are compared with Gallilei
invariant scattering calculations. Again the calculations were
preformed for  laboratory angles $\Theta = 15^\circ$ and $\Theta =
60^\circ$. The
Gallilei  invariant calculations differ significantly from the Lorentz
invariant results. We did calculations starting with correct
 relativistic energy (dashed line in Fig.\
\ref{Florentz2}), and correct relativistic  momentum (dashed dotted line
in Fig.\ \ref{Florentz2}). Both calculations deviate strongly
from the correct relativistic
calculation (solid line in Fig.\
\ref{Florentz2}).

To show that the two-nucleon
$\sqrt{s}$-distribution  used in the OSMC simulation and the
AMD agrees with the relativistic prescription we compared the
$\sqrt{s}$ distributions of the OSMC and the Lorentz invariant
scattering.
Again there are only
small differences  for  the resulting $\sqrt s$-distribution  as seen
in Fig.\ \ref{Florentz3}: OSMC
simulation (dashed  line), Lorentz invariant kinematics (solid line).
The averaged deviation of $\sqrt{s}$ is less than 1\% and, since the
elastic and inelastic nucleon-nucleon cross sections in the considered
energy region  have a smooth dependence on $\sqrt{s}$, the difference
between OSMC and purely Lorentz invariant kinematics is not important.

\section{Comparison with Experiment}
\label{exper}
As   first step to understand the employed delta dynamics in AMD we
have calculated the inelastic scattering of protons on light nuclei.
We have   compared the calculation with experimental data in which
both the delta
excitation and the quasielastic peak (QEP) are prominent.
In the following discussion we will see that behind the seemingly easy
to understand experimental data complicated and interesting processes
are hidden.

\subsection{One-step Monte Carlo calculations}
\label{monteexp}
First we want to discuss the results of the OSMC calculations.
The total
reaction probabilities for the (p,p$'$) reaction for $E_{\rm
lab}=800$ MeV incorporated in the OSMC are  shown in Fig.\
\ref{Fcrossratio}.  The ratio  between
inelastic to elastic cross section is approximately 2
to 3, which points out that inelastic channels are very important for this
energy region. Deviations from the total reaction probabilities in the
differential cross sections discussed in the following, are due to the
anisotropic cross sections used in the OSMC and due to kinematical
reasons.

In Fig.\
\ref{Fmonte} we show the OSMC calculations  (solid line) for the (p,p$'$)
reaction at $E_{\rm lab} = 800$ MeV for laboratory angles $\Theta =15^\circ,
30^\circ,40^\circ$ and  $\Theta =60^\circ$. In the experimental data
for $\Theta =15^\circ$ the QEP can be clearly seen  and the contribution
of  inelastic or multi-step scattering to the cross section   is
well separated. In Fig.\ \ref{Fmonte} we
show two different experimental data for  $\Theta =15^\circ$, which
differ only in the QEP region. This difference is claimed to be due to the
resolution of the experimental apparatus \cite{tan82}.
We scale  the  OSMC calculation to reproduce the
momentum region  p$'$=$600$-$800$ MeV at $\Theta =
15^\circ$. To achieve this, we have to use $X_{\rm factor} =260 $ mb
which is larger than  the measured total cross section (209 mb for
$E_{\rm lab} = 860$ MeV in Ref.\ \cite{che55} and 229 mb for $E_{\rm
lab}$ = 553 MeV in Ref.\ \cite{ren72} ).

We find that in the OSMC calculation the QEP for $\Theta =
15^\circ$ is largely overestimated and for $\Theta = 30^\circ$ and $\Theta
= 40^\circ$ the QEP
is much more pronounced than in the experimental data. The high
momentum tail  in the experimental data at $\Theta =15^\circ$ is physically
not allowed and consequently cannot be reproduced.

For   $\Theta =
30^\circ, 40^\circ$ and  $\Theta = 60^\circ$ there is an
underestimation of the low and high momentum tail of the cross section
in the shown momentum region.
The underestimation is especially strong in the momentum region  p$'
= 500$-$800$ MeV for $\Theta = 30^\circ$  and p$'
= 500$-$700$ MeV for $\Theta = 40^\circ$and . Another defect of the OSMC
calculation is that the flat behaviour of the cross section for p$'
=1000$-$1200$ MeV at $\Theta = 15^\circ$ is not reproduced. Instead
there is a dip in the
calculated cross section.

\subsection{AMD calculations}
\label{amdexp}
Before discussing the results of the AMD calculation we have to
comment on the cross section calculation in AMD for the (p,p$'$)
reaction. A straightforward approach leads to Eq.\
(\ref{crosscal}) for the double differential cross section. This
formula is based on the assumption that every
nucleon in AMD has the same width $\nu$, but this is not true for all
nucleons in the (p,p$'$) reaction. For example  it is
obvious that the incoming proton has a definite energy and therefore
zero momentum width $\nu$. Therefore the role of the width parameter
has to be reconsidered.
Concerning the width parameter $\nu$ in the AMD, we recall, that
the width parameter $\nu$ in the AMD is improtant for
reproducing the total
density and momentum distribution of ground state nuclei. In the
collision prescription in the
present version of the AMD  (see Ref.\
\cite{ONOA}), only the Gaussian distribution in the coordinate space is
taken into account, whereas  in the momentum space only the centers of
Gaussian wave packets are used.  Therefore in  nucleon-nucleus collision
calculation with AMD  the
momentum distribution of the target is only partly explored
dynamically. For heavy ion reactions the  effect of the high momentum
tail of the wave function is negligible  as
long as subthreshold processes are not important for the calculated
process, but, as will be
become clear   in the following, the QEP region of  the (p,p$'$)
reaction is sensitive to the
width parameter $\nu$ of the outgoing nucleons.

In order to solve the above sketched problem, for getting  a reliable
exploration
with AMD of  the  momentum distribution of the
target in the (p,p$'$) reaction,  two lines of argumentation  can be
followed.  One is to
take into account the momentum distribution  of the
colliding nucleons in the collision prescription, and the other is to
change the
width parameter $\nu$ of the outgoing nucleons.  In the first approach
 the width parameter $\nu$ of the incoming proton  should  be
zero, at least in the simulation of the nucleon-nulceon  collison.
Technically
 this can be taken into account in the collision prescription, but it
is not clear what
width should be attributed to the incoming proton after the first
collision.
We follow the  second approach to get a reliable treatment of the
momentum distribution of the target, since there is a way to
determine the width $\nu$ of the outgoing nucleons independently of
experimental data, namely, the AMD calculation should reproduce the
width of the QEP caused by the Fermi motion.  In course of
doing so we did calculations both with
the OSMC and a modified version of the  AMD, the 'one-step AMD'.  In case of
the OSMC we used both, a box momentum distribution for the target, like in
nuclear matter, and a Thomas Fermi momentum distribution for the
target, as explained
in Sec.\
\ref{montefo}.  In the 'one-step AMD' calculation we allowed only  one
possible collision in each event  after which the particles
propagated in their mean field.
We determine the width $\nu$ of the outgoing nucleon so that the
'one-step AMD' calculation reproduces the width of the QEP given by the OSMC
calculation with the Thomas Fermi momentum distribution for the
target.

  As can be seen in Fig\ \ref{Fmonte}, the QEP at
$\Theta =15^\circ$ is much narrower  than at $\Theta =30^\circ$. The
width parameter $\nu$ in the 'one-step AMD' calculation is fitted so
as to
reproduce  width of the QEP caused by the Fermi
motion. In the actual calculations  we did not
change the width parameter $\nu$ but we used a cut of the
Gaussian momentum distribution
\begin{equation}
\rho' (p) = 0 \quad {\rm for} \quad    \Bigl[  {1 \over 2 \hbar^2 \nu}
( {\bf p} -
{\bf K}_i )^2
  \Bigr] > Q_{\rm cut} \quad ,
\label{cut}
\end{equation}
and renormalized the function $\rho' (p)$ in Eq.\ \ref{crosscal}, which
has the same
effect as adjusting the
 width parameter $\nu$ in the single particle wave functions.

We found  that in the 'one-step AMD' calculation for $^{12}$C(p,p$'$)
reaction we have to use $Q_{\rm cut}=0.5$ for
$\Theta =15^\circ$, $Q_{\rm cut}=1.0$ for $\Theta =30^\circ$,
$Q_{\rm cut}=1.6$ for $\Theta =40^\circ$ and no cut for $\Theta
=60^\circ$ in order to
reproduce the width of the QEP in the OSMC simulation with Thomas
Fermi momentum distribution of the $^{12}$C target.

In all  AMD calculations we use
the above determined  cuts.

By comparing the 'one-step AMD' with OSMC calculations with box
momentum  distribution, the
nuclear matter case, we find that  we only
have to apply a cut at $\Theta
=15^\circ$ to reproduce the effect of the Fermi motion. This is a nice
result which states that for cross section calculations for heavy ion
collisions
no cuts are needed, since large amount of nuclear matter is involved
in the reaction.

After fixing all parameters we can proceed to discuss the results of
the AMD calculations which are shown in Fig.\ \ref{Famd2} (solid
line). In general we find a better reproduction of the experimental
data with the AMD than  with the OSMC. Here we will only state the
improvements established
by using the AMD instead
of the OSMC. In the following section (Sec.\ \ref{Sstep}) we study the
reaction mechanism
in the AMD calculation in detail and in Sec.\ \ref{Sdiff} we will
discuss the origin of the improvements in the  AMD.

Instead of an underestimation of the low momentum region for angles
$\Theta =30^\circ$ and $\Theta =40^\circ$ in the OSMC, we find that
the experimental data can be better reproduced in the AMD
calculation. In general the agreement with the
experimental data  for these angles  is much better in the AMD
than in OSMC,
because the
smooth behaviour of the experimental data for  $\Theta =30^\circ$, and
especially for  $\Theta =40^\circ$, is much better reproduced. In the
case of the OSMC we do not get absolute cross sections and therefore
the comparison of the calculations with  the shape of the momentum
(p$'$) dependence is important
and this is much better reproduced in the AMD calculations.

The AMD can also reproduce the  high momentum tail of the cross
section for $\Theta =30^\circ$ better and for $\Theta =40^\circ$ and
 for  $\Theta =60^\circ$ much better than the OSMC. Also the
 the flat behaviour of the cross section for p$'
=1000$-$1200$ MeV at $\Theta = 15^\circ$ is  reproduced.

So we find that the agreement between AMD calculation and experiment
is much better than the one between OSMC   and experiment. The
experimental data can be quantatively reproduced in AMD, whereby
the elastic processes  seem to be overestimated in forward
direction. In Sec.\ \ref{experpara} we will discuss how to improve the
AMD results, but in the next sections we will first try to get a
clearer understanding of the reaction dynamics in the AMD for the
(p,p$'$) reaction.

\subsection{Multi-step contributions to the cross section}
\label{Sstep}
To get a better understanding of the reaction dynamics we calculated
the decomposition of the cross section into different multi-step
contributions and reaction processes. The result of this decomposition
is shown in Fig.\ \ref{Fstepcont}. We show calculations for four
angles: $\Theta =15^\circ, 30^\circ, 40^\circ$ and $\Theta = 60^\circ$. The
elastic multi-step contributions shown in Fig.\ \ref{Fstepcont}
 are protons with  one-step
elastic scattering
(ELA-1step, long dashed  line) and protons with two-setp elastic scattering
 (ELA-2step,
 dashed  line). Nucleons involved in inelastic collisions can contribute
 to two different sources of outgoing protons: protons
emitted as delta decay products and protons  which
acted as collision partners in the delta excitation process. The
contributions of these different sources in inelastic collisions are
also shown in Fig.\ \ref{Fstepcont}: Protons emitted as partner of an
excited delta in
the first chance collision ($\Delta$-1step(N), dotted line), and protons from
delta decay after the first chance collision($\Delta$-1step($\Delta$),
dashed dotted line).

In the calculation for $\Theta =15^\circ$ we see that mainly  the one-step
scattering process (long dashed line) contributes to the QEP. The two
step contribution
 is much smaller for all momenta. A similar  behaviour is found  for
$\Theta =30^\circ$, where the two step contribution (dashed line) gets
more important
but  is smaller than the one-step contribution  for almost all
outgoing proton momenta. At $\Theta = 30^\circ$, $\Theta = 40^\circ$
and $\Theta =60^\circ$ the two
step contribution dominate the high momentum tail of the proton
spectrum, since
 one-step scattering cannot reach this region of the
phase space easily.

For $\Theta =30^\circ$ and $\Theta =40^\circ$ we find that for the
lower momentum part of the cross section, below the QEP, many higher
step processes contribute to the cross section. Besides the
processes shown in Fig.\ \ref{Fstepcont} two-step and three-step
elastic processes followed by an inelastic process  contribute to the cross
section. Therefore the reproduction of the smooth behaviour of the
cross section in this momentum region is quite sensitive to the ratio
of inelastic to elastic cross section and the multi-step contributions.

\subsection{Comparison of AMD and OSMC: Multi-step and potential effects}
\label{Sdiff}
After having discussed the reaction dynamics in the AMD in detail we
can comment on the differences between AMD and OSMC and what kind of
physical effects  are important for the (p,p$'$) reaction.

As one could have expected, the main differences of OSMC and AMD are due
to multi-step processes, but the details of the origins of the
differences are much more interesting than this plain statement, and
therefore we discuss this question in detail. Also we have to
investigate the role of potential effects which could interfere with the
multi-step contribution effects.
To make the above statements more transparent we show in Fig\
\ref{Fonenewd} the
outcome of the 'one-step AMD' calculation (dashed line) we mentioned in Sec.\
\ref{amdexp}, in which we did not allow anymore collisions after
the first proton-nucleon collision ocurred in the AMD, propagating the
particles from this time step onwards only in their
mean fields without collisons.
Therefore the only difference between 'one-step AMD' and OSMC is the
treatmant of the mean fields in the AMD and the differences between
'one-step AMD' and AMD are only due to multi-step contributions. The results
of the 'one-step AMD' should not be confused with the one-step
contributions to the full AMD calculations discussed in the last
section.

First we want to discuss the differences due to multi-step
contributions and the improvements achieved by this in reproducing the
experimental data with the AMD. Therefore we compare 'one-step AMD'
(dashed  line) and AMD
(solid line) calculations in Fig.\ \ref{Fonenewd} on the basis of the
 multi-step decompositon  of the AMD results shown in Fig.\ \ref{Fstepcont}.
 Multi-step contributions which involve at least one inelastic
step are mainly responsible for a
better reproduction of the experimental data for momentum below the
QEP at $\Theta = 30^\circ$ and $\Theta = 40^\circ$ in the AMD
calculation. The high momentum
tails above the QEP at ($\Theta = 30^\circ$,)  $\Theta = 40^\circ$ and
$\Theta = 60^\circ$ are better reproduced in the AMD by mainly
multi-step elastic scattering.

The apparent quantitative agreement
of the heights of the QEPs for   $\Theta = 30^\circ$  and $\Theta =
40^\circ$
 in the 'one-step AMD' and the AMD is not a trivial result. The
reduction of the QEP due to multi-step scattering is balanced out by an
enhancement due to multi-step processes in the AMD.  At $\Theta =
60^\circ$ multi-step processes lead even to an increase of the cross
section in the QEP region in the AMD calculations. For $\Theta =
15^\circ$ the net effect of multi-step contributions is a reduction of
the QEP.

Another important improvement of the AMD calculations due to
multi-step contributions is that the flat behaviour of the
experimental data at $\Theta =
15^\circ$  for momenta p$'$= 1000-1200 MeV can be described unlike
in the OSMC
where there is a dip in the cross section.

Concerning the potential effects in AMD we find no significant
difference between 'one-step AMD' and OSMC calculations. There are no
 apparent  differences caused by the inclusion of potentials for the
particles. (See more detailed discussion about delta potential effects
in the next section.)

\subsection{Dependence on delta potential, cross section and target momentum
distribution, and improvements}
\label{experpara}
Since we have now established a detailed understanding of the
reaction dynamics in the AMD we can study the dependence of the
results on the  physical
input quantities and what we can learn about these in the (p,p$'$)
reaction.

One important question which is always asked, but not yet well understood, is
the question of the delta potential and the behaviour of the delta in
the nuclear  medium. We want to comment on this from the viewpoint of
the AMD calculations made
in this paper.  In the previous section we stated that there is no
apparent potential effect when comparing OSMC and AMD results, but
here we will investigate the potential effects in more detail.

To study the effect of the delta
potential we did calculation with a potential which has the same
density dependence as the one
in Eq.\ \ref{DELTPOT}, but is shifted to reproduce  $U_\Delta (\rho_0) =
-80$ MeV. This choice of the potential is motivated by the following
idea: We assume that the delta potential is momentum-independent
 and has  the
original value $U_\Delta (\rho_0) = -30$ MeV. We recall here that we
have not  treated the
momentum dependence of the nucleon potential correctly in the present
version of the AMD and our nucleon potential at $E_{\rm lab} =800$ MeV
is $U_N (\rho_0) \approx 0$ MeV. If we used the correct value $U_N
(\rho_0) \approx 50$ MeV, it would
lead to the difference $U_N (\rho_0)-U_\Delta (\rho_0) \approx  80$
MeV. Hence using
a delta potential with  $U_\Delta (\rho_0) =
-80$ MeV in the present version of the AMD leads to the desired value
$U_N (\rho_0)-U_\Delta (\rho_0) \approx  80$ MeV.

The AMD calculations with shifted delta potential are shown in
Fig.\ \ref{Fonenewd} (dotted line).
There is no apparent difference between the calculations with
different delta potentials. Only if we calculate the multi-step
decomposition of the cross section we find that the high momentum
threshold of the one-step inelastic contributions are different for
different delta potentials. But these differences cannot be seen in
the sum of all contributions  because in the high momentum threshold
region for these processes their contributions to the cross sections are
just a few mb and therefore are hidden below the multi-step and elastic step
contributions to the cross section. This is also the reason why we did
not find any apparent potential effect in the previous section.

We point out at this point that the choice of the delta mass in
the $NN\rightarrow N\Delta$ process  is not treated fully self consistent in
the present version of the AMD. To determine the mass of the delta we
use the delta mass distribution Eq.\ \ref{massdist} dependent on the
$\sqrt{s}$ of the initial nucleon-nucleon pair, but we do not take
the
effect of the delta potential on the delta mass distribution into
account. This improvement should be made in future  AMD
calculations. Still the above
findings about the influence of the delta potential on AMD
calculations for experimental data for the  $^{12}$C(p,p$'$) reaction
 would hold. This is due to the fact that in the inelastic
threshold
 region (see argumentation above) where the change of the delta
potential has the
biggest   effect, the multi-step and elastic scattering
dominates the cross section. To study the influence of the delta
potential we have to compare with experimental data which are  mainly
dominated by the excitation of deltas, i.e.\ delta mass distribution
from $\pi$ + nucleon invariant mass spectra (see for example
Ref.\ \cite{chi91}) in (p,x)x$'\pi$ reactions.

Concerning the dependence on the angular distribution of the adopted
elastic nucleon-nucleon cross section we find strong dependence on the adopted
parametrisation.
Since the OSMC calculation is much less time consuming we  studied
the cross section dependence within the OSMC.
 In Fig.\ \ref{Fbuuiso} we show the result of the OSMC calculation for
isotropic
elastic nucleon-nucleon cross section (solid line) and for the
isospin-independent
Cugnon parametrisation \cite{cug81} for the elastic nucleon-nucleon cross
section. Since the isospin-independent Cugnon parametrisation is
strongly forward peaked for all nucleon-nucleon channels, at $\Theta =
15^\circ$ the
QEP is largely overestimated and at larger angles the QEP is
underestimated (Fig.\ \ref{Fbuuiso} (dashed line)). In case of an
isotropic nucleon-nucleon cross section the opposite is the case: The QEP
at  $\Theta =
15^\circ$ is  much better reproduced, but for all other angles
there is a large overestimation of the QEP (Fig.\ \ref{Fbuuiso} (solid
line)).

If we would do AMD calculations with  the isotropic or the
isospin-independent  parametrisation of the nucleon-nucleon cross
section, neither
choice would  improve the results, which we achieved
using the isospin-dependent parametrisation of Cugnon. This is a
satisfying result, since the    isospin-dependent parametrisation of
Cugnon gives the best
reproduction of the experimental data  among the dicussed
parametrisations.

The strong dependence on the angular distribution of the
nucleon-nucleon cross section and the effect of multi-step scattering
discussed above show how difficult it is to explain the considered
experimental data for the (p,p$'$) reaction which at first view seem
to be very dull,
uninteresting and easy to understand.

With the OSMC we  checked the dependence of the results on the
adopted  momentum distribution of the target. In Fig.\
\ref{Fmonte}, besides the original OSMC calculation (solid line), also
calculations with a box momentum distribution of the target (dotted
line)  are shown. The higher momentum components in the target lead to
a broadening
 and reduction of the QEP. But in general the (p,p$'$) reaction is not
 sensitve to the details of the momentum distribution in the
target. The large variation of width of the QEP between Thomas Fermi
 and box
 momentum distributions is caused by a large difference of the
distributions themselves.

We did calculations with different mass distributions for the
delta discussed in Sec.\ \ref{amdcoll}, but these only have a minor
effect on the result of the calculation.

We  have checked the effect of Pauli blocking in the OSMC. All
OSMC calculations presented in this paper incorporate Pauli
blocking. In the AMD the Pauli blocking is incorporated naturally (see
Ref.\ \cite{ONOA} for details). We found that only the high momentum tail of
protons at  $\Theta =15^\circ$
is influenced by the consideration of Pauli blocking.

In general we see in  the multi-step decomposition of the AMD
calculation in Fig.\ \ref{Fstepcont} that the
contribution of elastic scattering is overestimated. This is especially
prominent for $\Theta =30^\circ$ but also for  other angles the defect
can be  seen. As a first attempt to cure this problem we did calculations in
which we reduced the elastic nucleon-nucleon cross section
artifically by
50\%. This leads to the results shown in Fig.\
\ref{Famd2} (dashed line). The QEP can be reproduced much better and also
the general agreement between  experimental data and AMD calculation
gets  better. At $\Theta =40^\circ$ the smooth momentum dependence of
the cross section is well described but the cross section is slightly
underestimated.

The reduction of the elastic cross section is also motivated by the
fact that we
find a  total reaction cross section of 307 mb for the
$^{12}$C(p,p$'$) reaction  at $E_{\rm lab}=800$ MeV
in the AMD calculation, which is too large compared to the
experimental results (see Sec.\ \ref{monteexp}).

With the OSMC we did calculations with a 70\% reduced elastic cross
section. For this choice of the cross section we have to use $X_{\rm
factor}=190$ mb which is less than  the  measured total cross
section.   The results are shown in Fig.\ \ref{Fmonte} (dashed
line). Also in case of the OSMC the reduction of the cross section
leads to a better agreement with  the experimental data
for $\Theta =15^\circ$ and  $\Theta =30^\circ$.
 But  in the  case $\Theta =40^\circ$
and  $\Theta =60^\circ$ the agreement with the experimental data gets
worse.

One possible physical effect which will lead to  a
reduction of elastic scattering events is the correct treatment of the
momentum dependence of the mean field. As written above the adopted
Gogny force does not give the correct momentum dependence of the
nucleon-mean potential. Instead of $U_{N}(800 {\rm MeV})\approx
50$ MeV the Gogny force leads to  $V_{N}(800 {\rm MeV})\approx
0$ MeV. Using a repulsive
potential would lead to more nucleons diffracted to larger angles. This
also implies that nucleons would have less interactions with the
target since they are diffracted from their way through the
target. This effect is  especially true for  nucleons with large
impact parameter which contribute strongly to the elastic scattering
peak. In a future investigation this idea should be tested.

Finally we shortly want to comment on the differences of the AMD
 and OSMC calculations for the $^{12}$C(p,p$'$) reaction at $E_{\rm
lab}=800$ MeV
from the PWIA calculations by Y.\ Alexander et
al.\ \cite{ale80} for the same process. Y.\ Alexander et al.\ take
one-step elastic and
one-step inelastic scattering into account. Unlike  in the OSMC
results discussed in this paper the smooth dependence on the outgoing
proton momenta for larger angles can be better described in their PWIA
calculation. This is because they use an unrealistic high pion
production cross section.
 Therefore they need no multi-step contributions to get the smooth
behaviour of the cross section for large angles as we find in
the AMD calculations.  If one would add the
multi-step contributions to their PWIA calculation this would lead to a
large overestimation of
 the experimental data.

Another defect of the PWIA calculations of Y.\ Alexander et al.\ is that
 they fit the heights of the QEP in the double differential
cross section with an angle dependent scale factor which hides the
strong dependence on the angular distribution of the elementary
nucleon-nucleon cross section.

In the region of phase space where multi-step contributions
determine the cross section Y.\ Alexander et al.\  underestimate the
inelastic scattering cross section.
This is in good agreement with the findings presented in this paper
using the OSMC
model.

\section{Summary}
\label{summ}
In this paper we discussed how to incorporate delta degrees of freedom
into the AMD. Especially we discussed the inclusion of inelastic
channels in the collision term and the form of the delta-nucleon
potential. For the delta-nucleon potential we used a Skyrme type
potential which gives $U_{\Delta}(\rho_0)\approx -30$MeV for infinite nuclear
matter calculation, thereby reproducing the experimental data found in
pion-nucleus scattering.

With the help of the frictional cooling method we calculated
delta separation energies for 'delta nuclei'. There we found that the
minimum energy of the delta-nucleus system in the present calculation is
achieved  when the delta and nucleus wave functions are in the  state
of maximum overlap. Therefore we found that the clustering of AMD
wave function is reflected in the delta separation energy.

For a detailed discussion of the (p,p$'$) reaction we developed a
one-step Monte Carlo (OSMC) model for this reaction. As required by the
energy region considered in this paper ($E_{\rm lab} \approx 800$ MeV)
we incorporated both nucleon and delta degrees of freedom using the same
cross section as done in the AMD.

As first application of the OSMC we investigated the question of
Lorentz invariance
for  nucleon-nucleus scattering for energies $E_{\rm lab}
\approx 800$ MeV. We found the interesting result that, applying a
combination of
Lorentz transformations, also a Gallilei invariant theory can be used
to treat nucleon-nucleus scattering in this energy region. Also the
calculation of $\sqrt{s}$ in a
Gallilei invariant theory can be performed in an approximate Lorentz
invariant way.

Before applying the AMD calculation to the (p,p$'$) reaction, we
discussed how to determine the width parmeter $\nu$ of the single
particle  wave function for outgoing nucleons. The determination was
made  by comparing 'one-step
AMD' and OSMC calculations. This is a unique way to get parameter
independent results in the AMD.

Both, with extended AMD and OSMC, we preformed calculations of double
differential cross sections $d\sigma^2/dp'/d\Omega$ for the
$^{12}$C(p,p$'$) reaction for angle and momentum-dependent cross sections. We
found a quantitative difference between AMD and OSMC calculations which
is mainly due to  multi-step contributions to the cross
section, which are absent in one-step calculations.

Generally we find that OSMC and AMD can reproduce the qualitative
behaviour of the experimental data. Both the delta peak and the QEP can be
seen in the calculation and the angle dependence of the cross section
is reproduced rather
well. In case of the AMD we find a quantitative  good description of the
experimental data expect for the overestimation of the QEP in forward
direction.
 Therefore the agreement with
the experimental data improves  if we perform calculations with a
reduced elastic cross section.
One reason for the reduced elastic cross section could be that the
correct treatment of the momentum dependence of the  nucleon mean
field
at this energy
region leads to a repulsive potential and, because of this, to  more
diffraction of the incoming protons to larger angles. This then leads
to less elastic scattering
probability. This effect would be most serious for big impact
parameters.

For a better understanding of the reaction process we preformed a
multi-step decomposition of the cross section. To get the smooth momentum
dependence seen in the experimental data for $\Theta =30^\circ$ and $\Theta
=40^\circ$, a complicated  interplay of elastic and inelastic channels is
important which is approximately reproduced by the AMD
calculation.

One interesting result  of the multi-step
decomposition of the cross section
is that two and higher step contributions govern the high momentum
tail of cross sections for larger angles. The unsatisfying  point is that an
overall reduction of the elastic scattering cross section, which seems to be
needed to reproduce  the experimental data better, leads to an
underestimation of multi-step processes and therefore to an
underestimation of the cross section in the region of phace space
where these processes determine the
cross section, such as the region of the high momentum tail at larger
angles.  Again the above offered cure to the overestimation of
the elastic scattering, a correct treatment of the momentum dependence
of the mean field, would mainly reduce the one-step contributions and
not the multi-step contributions.

As one important physical property, which is often discussed but not
well understood, we discussed the influence of the delta potential on
the (p,p$'$) reaction. We found that the results for the (p,p$'$) are
not very sensitive to the choice of the delta potential. The reason
for this is that the contributions to the cross section which change
due to different  delta potentials are small and are hidden below
multi-step and elastic step contributions to the cross section. We
pointed out that the dependence on the delta potential will be more
apparent in calculations for experimental data, where the excitation
of the delta
dominates the reaction, i.e.\ delta mass distribution
from $\pi$ + nucleon invariant mass spectra (see for example
Ref.\ \cite{chi91}).

Commenting on prevoius calculations we found that the
AMD model gives a much more realistic picture of the $^{12}$C(p,p$'$)
reaction at $E_{\rm lab} = 800$ MeV than the  PWIA calculations of
Y.\ Alexander et al.\  in Ref.\ \cite{ale80} for the same process.

Finally we conclude with an outlook:
This paper leaves many open questions and ways to extend the field of
study. First of all a nucleon-nucleon potential should be adopted
which gives the correct momentum dependence of the nucleon mean
field and then the statements made about the change of the  heights of the
QEP should be
tested.

Also the full pion dynamics should be incorporated in the AMD, which
is expected to have a minor effect on the calculations presented in
this paper where we
treat a light target $^{12}$C,
since we find that the $\Delta N\rightarrow NN$ process is
negligble.
But for the study of the (p,p$'$) reaction
for heavier targets the correct treatment of pion absorption ($\Delta
N \rightarrow NN$
reaction)  and pion reabsorption ($\pi N \rightarrow \Delta$) will
be much more important.

With fully incorporated pion dynamics detailed studies of pion
production and pion-nucleus reactions can be done.

Another way to follow is to  perform a more detailed study of
dependence on delta
and nucleon potential. Especially a momentum-dependent
delta potential could be introduced and the difference between
momentun-dependent
and momentum-independent potentials could be discussed. For this
investigation  we have to compare to experimental data which are more
sensitive to the adopted
potentials. Also we have to introduce a self consistent treatment of
the delta mass in the AMD
for the the $NN\rightarrow N\Delta$ process as discussed in
Sec.\ \ref{experpara}.

Further, (p,n) reactions should be investigated in detail. For  this study  the
the difference of transverse and longitudonal delta
excition should and can be taken into account in the cross section. This will
give a nice tool to study the shift of the delta peak in the
(p,n) reaction.

\acknowledgements
One of us (A. E) wants to thank the Japan Society for Promotion of
Science for financial support for this project. Most of the
calculations for this research project  were performed with the
Fujitsu VPP500 of  RIKEN, Japan.

\widetext
\begin{figure}
\caption{Basic Feynman diagram for pion production in nucleon-nucleon
collisions.\label{Fnnnnpi}
}
\end{figure}
\narrowtext

\widetext
\begin{figure}
\caption{Delta binding energies for various nuclei calculated in AMD
with
frictional cooling method.\label{Fdeltbind}
}
\end{figure}
\narrowtext

\widetext
\begin{figure}
\caption{OSMC of the p(p,n)p$\pi^+$ process at $E_{\rm
lab}=830$ MeV with
different delta mass distributions. Experimental data are from
Ref.\ \protect\cite{chi91}. (See text for details.)\label{FpHn}
}
\end{figure}
\narrowtext

\widetext
\begin{figure}
\caption{The total reaction probabilities in the OSMC
simulation which are
determined by the  total cross sections for proton incident energy
$E_{\rm lab} = 800 $MeV.
 \label{Fcrossratio}}
\end{figure}
\narrowtext

\widetext
\begin{figure}[htb]
\caption{One-step  scattering  simulation for the quasi elastic process
for $\Theta =15^\circ$ and
$\Theta =60^\circ$ at $E_{\rm lab}=800$ MeV with full Lorentz invariant
kinematics (solid line)  and combination of Gallilei and Lorentz
invariant kinematics (OSMC) (dashed line).
(See text for details.) \label{Florentz1}
}
\end{figure}
\narrowtext

\widetext
\begin{figure}
\caption{One-step scattering   simulation for the quasi elastic
process for $\Theta =15^\circ$ and
$\Theta =60^\circ$ at $E_{\rm lab}=800$ MeV with full Lorentz invariant
kinematics (solid line)  and Gallilei
invariant kinematics starting with correct energy (dashed dotted line)
and correct momentum (dashed line). (See text for details.)
\label{Florentz2}
}
\end{figure}
\narrowtext

\widetext
\begin{figure}
\caption{The distribution of \protect$\protect\sqrt{s}$
in one-step scattering calculations
 at $E_{\rm lab}=800$ MeV with full Lorentz invariant
kinematics (solid line)  and combination of Gallilei and Lor\-entz
in\-vari\-ant kinematics (OSMC;  dashed line).
(See text for details.) \label{Florentz3}}
\end{figure}
\narrowtext

\widetext
\begin{figure}
\caption{OSMC calculations described in
Sec.\ \protect\ref{monte}
in comparison with experimental data for $d^2\sigma/d\Omega/dp'$ in
$^{12}$C(p,p$'$) reactions
for various angles $\Theta = 15^\circ, 30^\circ,
40^\circ$ and $60^\circ$ in the
laboratory system and $E_{\rm lab}=800$ MeV. Calculations are for
free elastic nucleon-nucleon cross section (solid line), 70\%
reduced elastic nucleon-nucleon  cross
section (dashed line) and
box momentum distribution for the target (dotted line). For the solid
and dashed lines the Thomas Fermi momentum distribution for the
target is used, and for the dotted line free elastic nucleon-nucleon
cross section is used. (See text for
details.)  \label{Fmonte}}
\end{figure}
\narrowtext

\widetext
\begin{figure}[thb]
\caption{AMD calculation with free elastic nucleon-nucleon cross
section (solid line)
and 50\% reduced elastic nucleon-nucleon cross section (dashed line)
in comparison
with experimental data for $d^2\sigma/d\Omega/dp'$ in  $^{12}$C(p,
p$'$) reactions at $E_{\rm lab}=800$ MeV. Experimental
data are the same as in Fig.\ \protect\ref{Fmonte}. \label{Famd2}}
\end{figure}
\narrowtext

\widetext
\begin{figure}[tbh]
\caption{Multi-step decomposition of the original AMD calculation
(solid line) given  in
Fig.\ \protect\ref{Famd2}. (See text for details). Experimental
data are the same as in Fig.\ \protect\ref{Fmonte}.
\label{Fstepcont}}
\end{figure}
\narrowtext

\widetext
\begin{figure}[htb]
\caption{Original AMD calculation (solid line), 'one-step AMD'
calculation (dashed line) and AMD calculation with shifted delta
potential (dotted line) in comparison
with experimental data for $d^2\sigma/d\Omega/dp'$ in  $^{12}$C(p,
p$'$) reactions at $E_{\rm lab}=800$ MeV. (See text for
details). Experimental
data are the same as in Fig.\ \protect\ref{Fmonte}.
\label{Fonenewd}}
\end{figure}
\narrowtext

\widetext
\begin{figure}[thb]
\caption{OSMC calculation with isotropic nucleon-nucleon cross section
(solid line), forward peaked nucleon-nucleon cross section (dashed
line) and box momentum distribution for the target (dotted line)
 in comparison
with experimental data for $d^2\sigma/d\Omega/dp'$ in  $^{12}$C(p,
p$'$) reactions at $E_{\rm lab}=800$ MeV. (See text for
details). Experimental
data are the same as in Fig.\ \protect\ref{Fmonte}.
\label{Fbuuiso}}
\end{figure}
\narrowtext

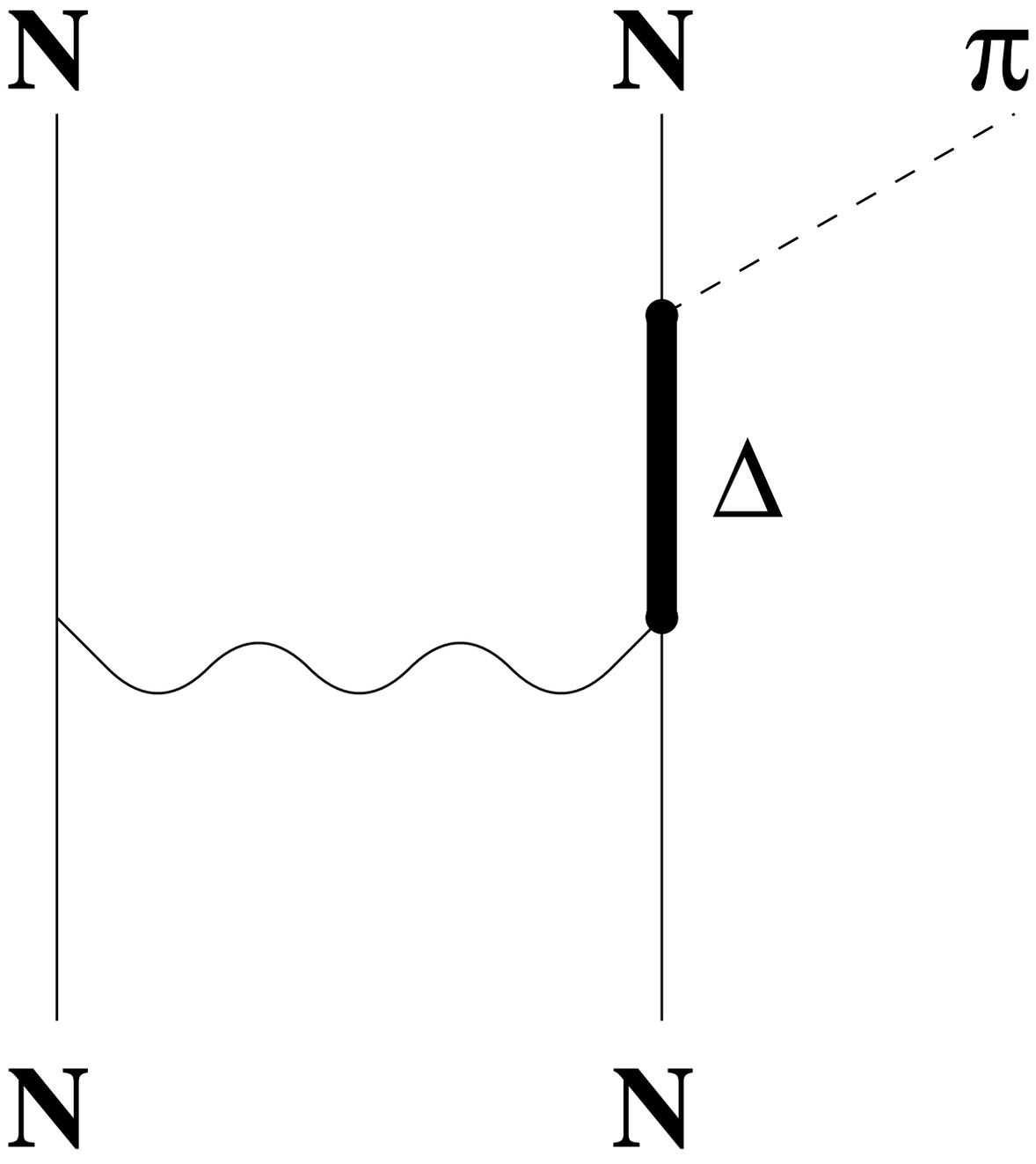
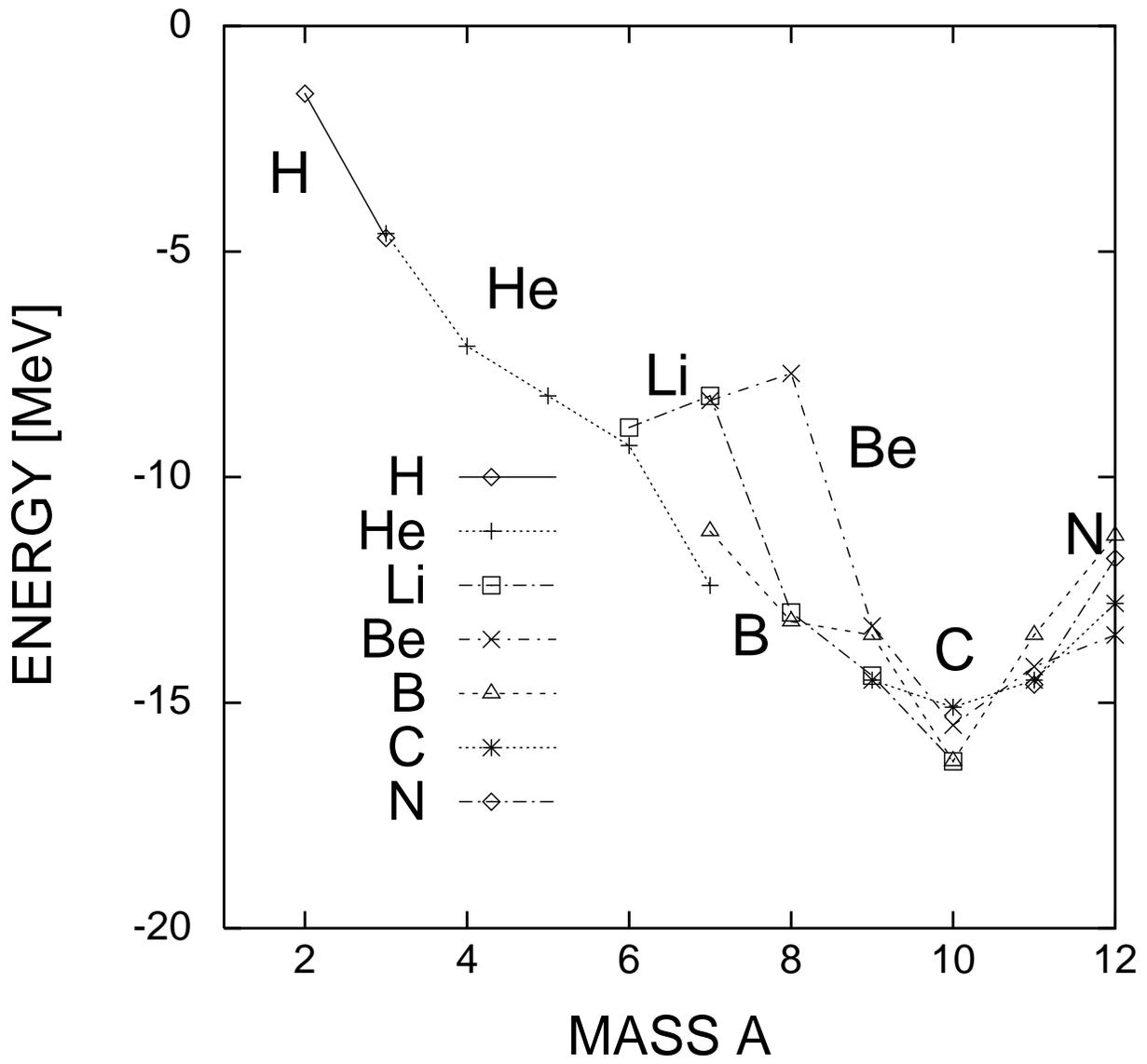
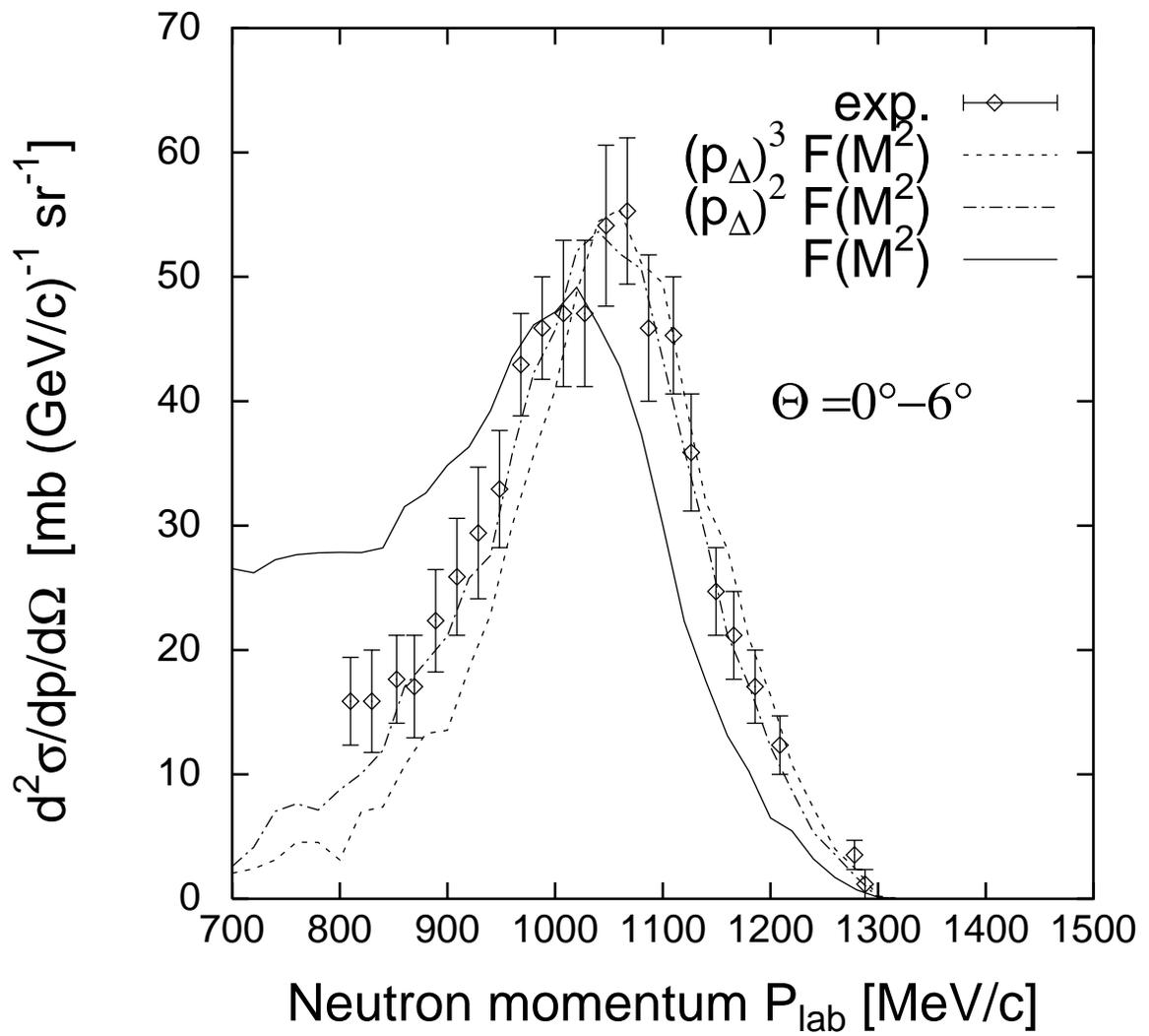
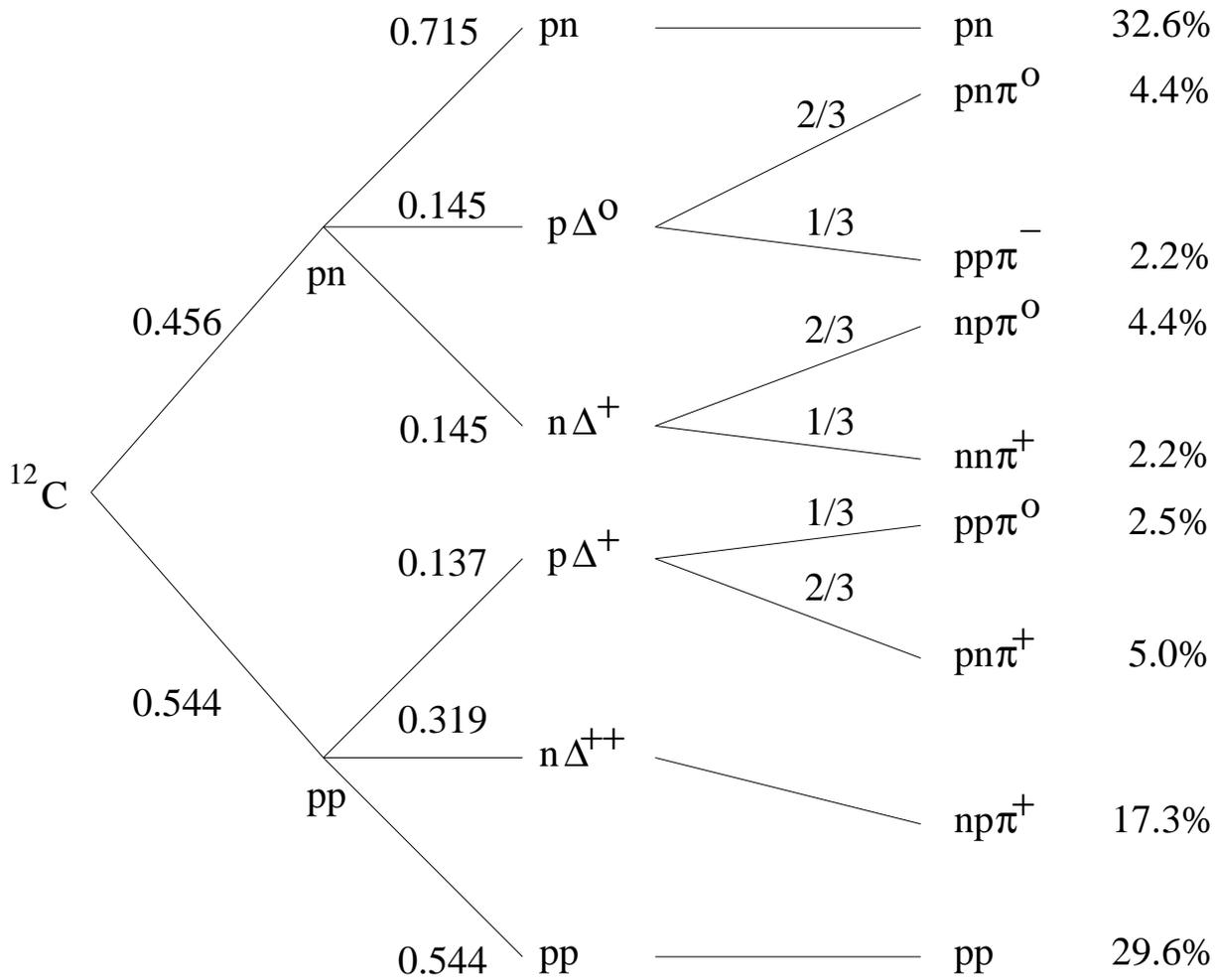
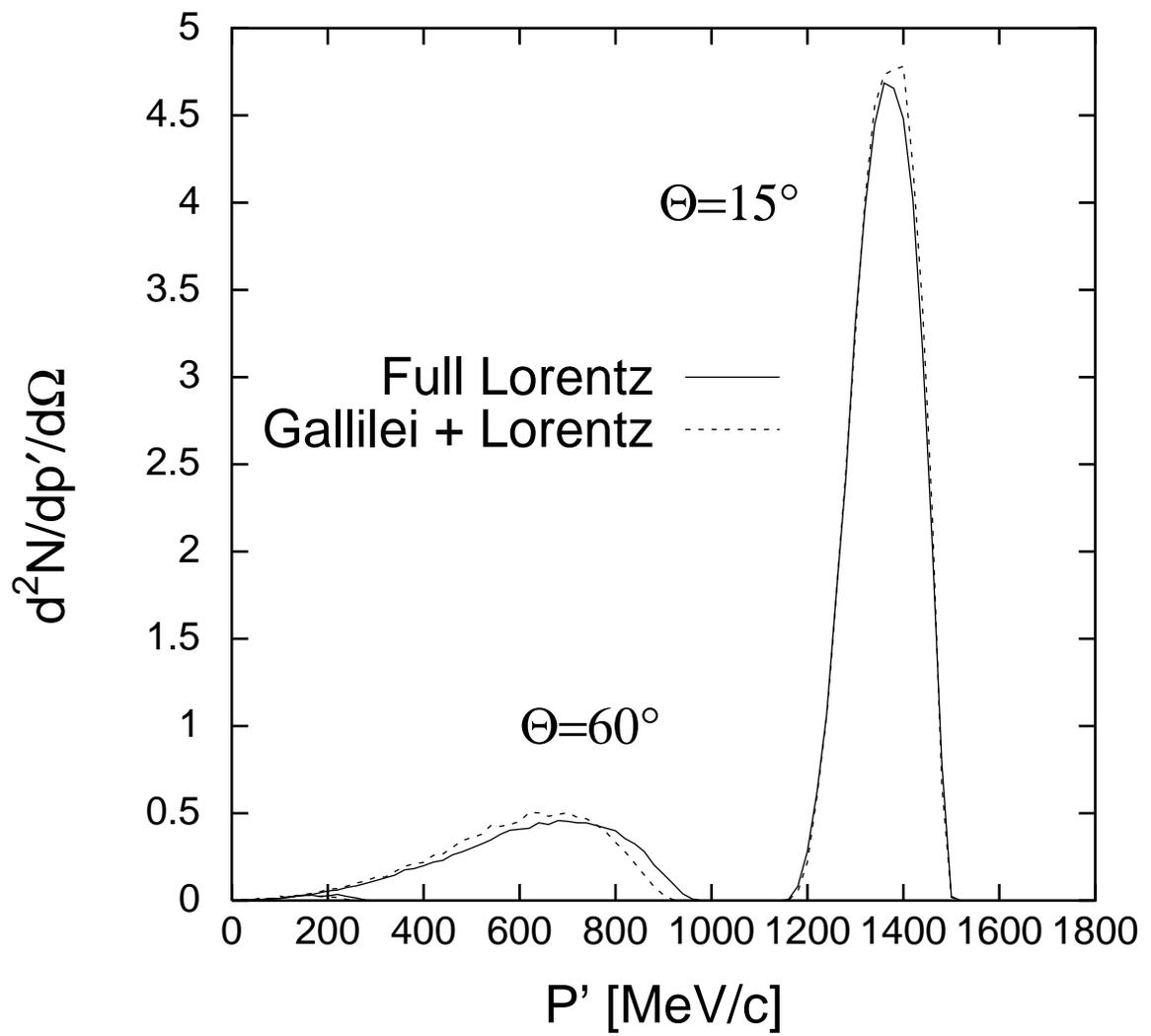
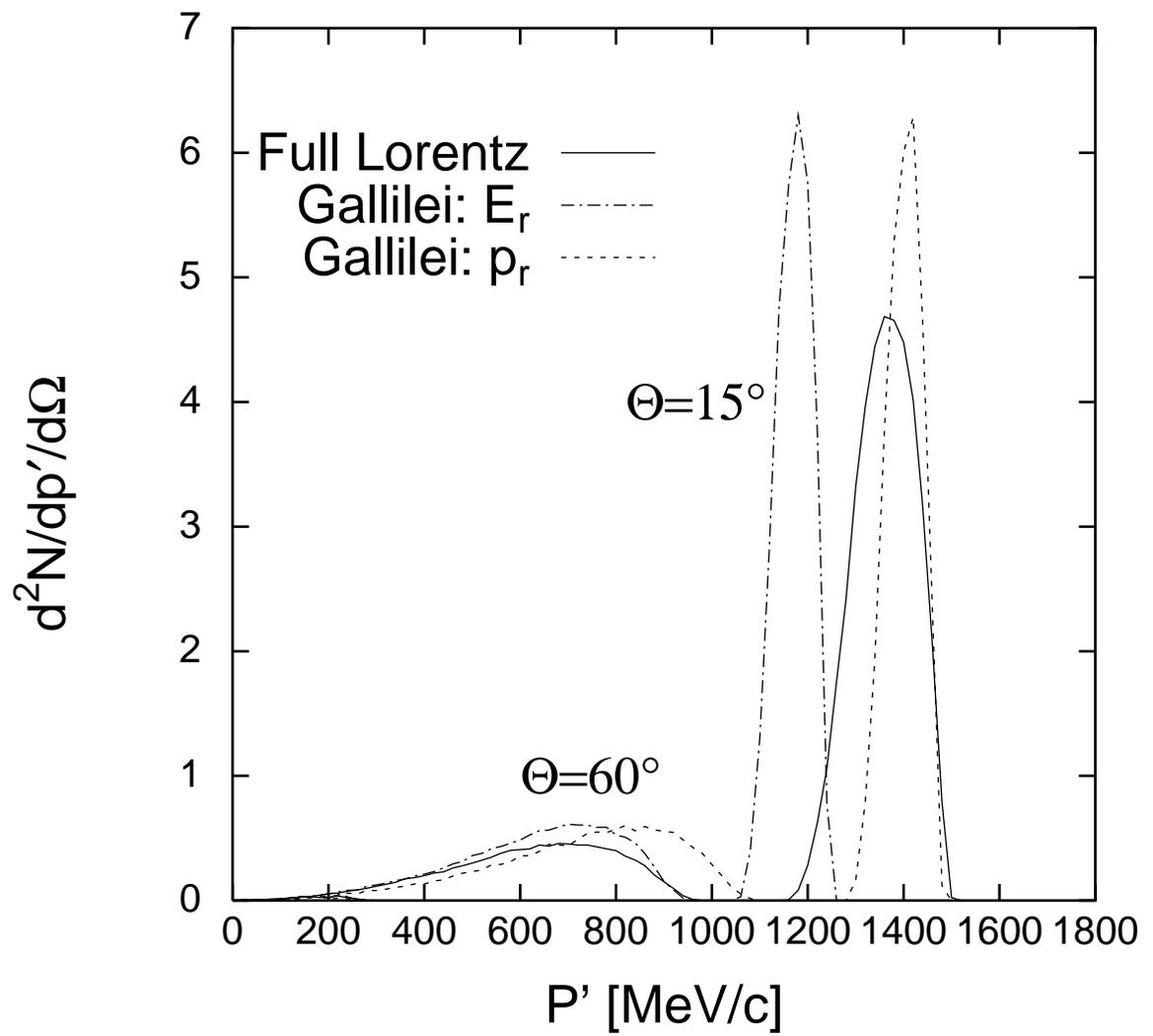
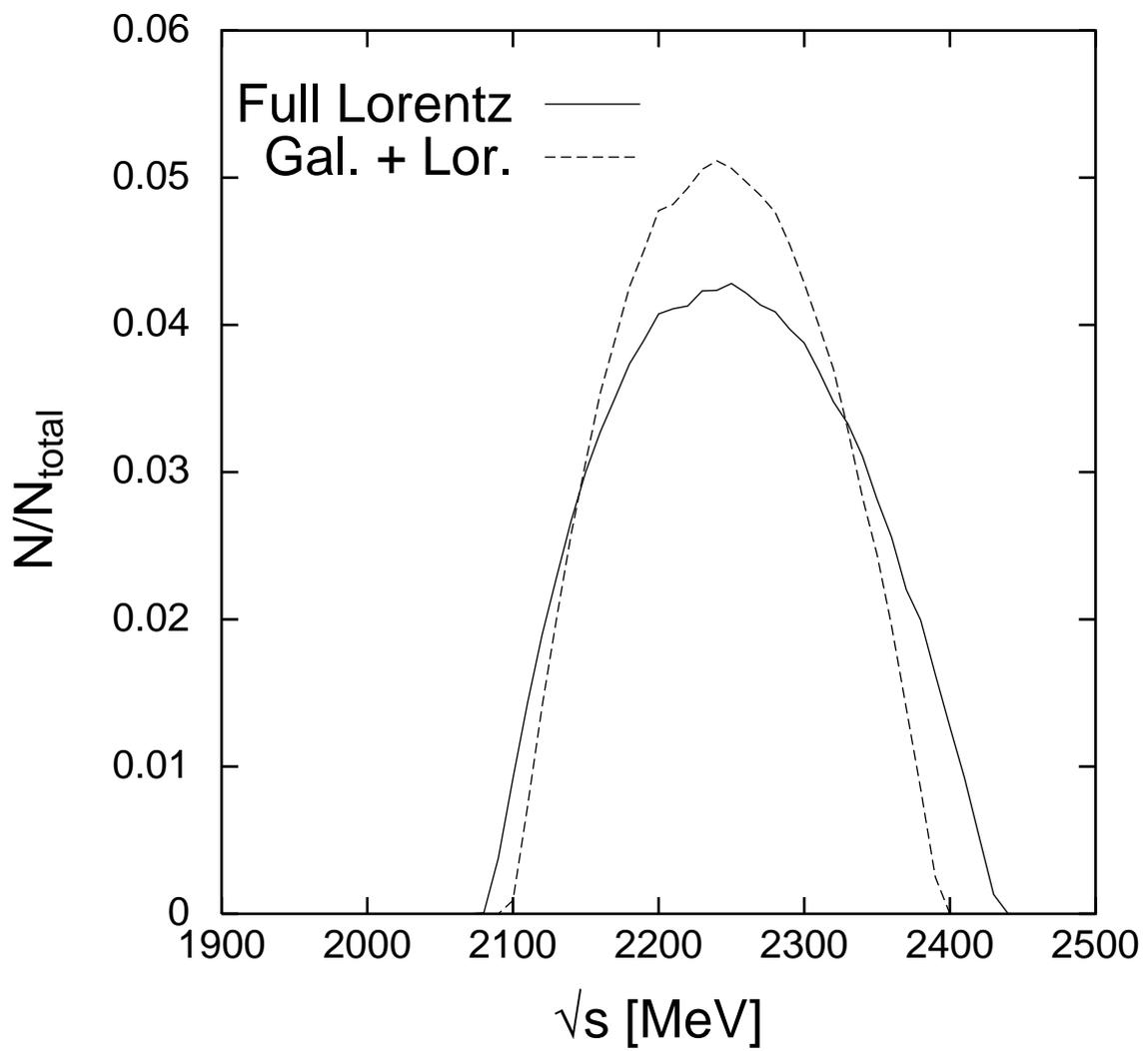
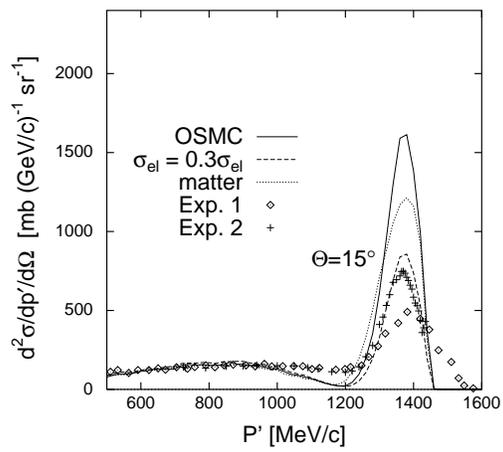
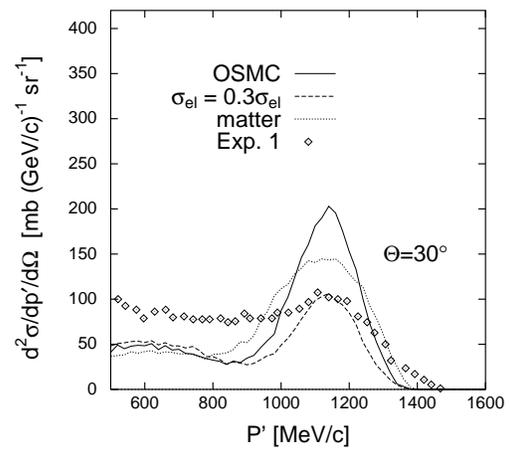
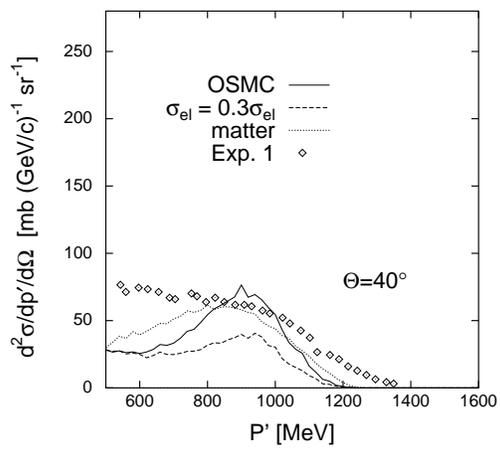
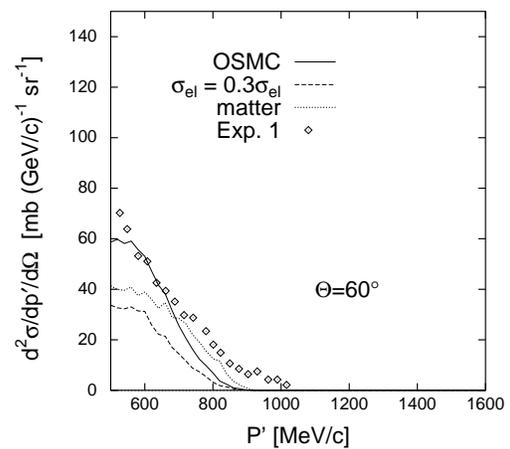
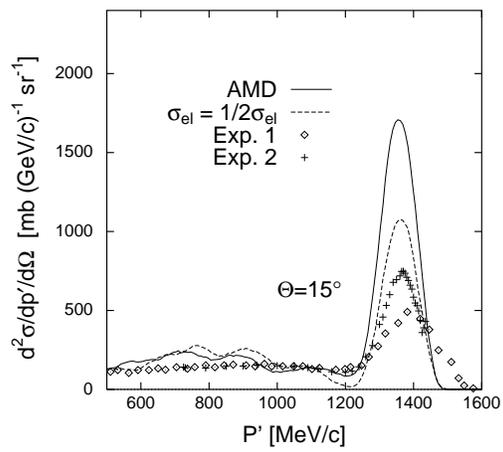
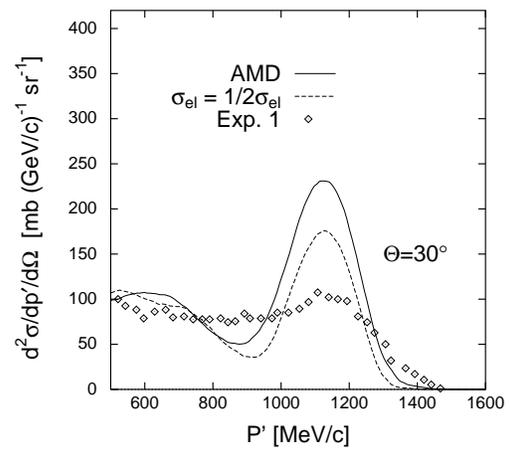
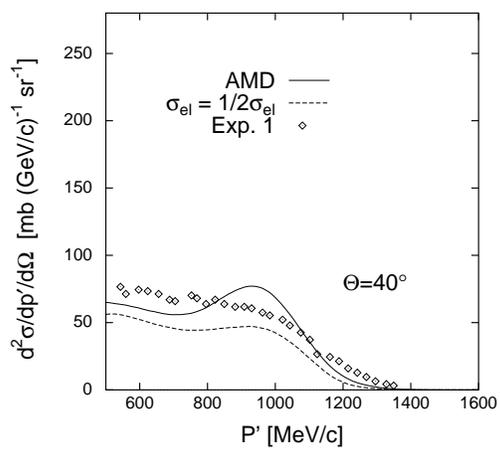
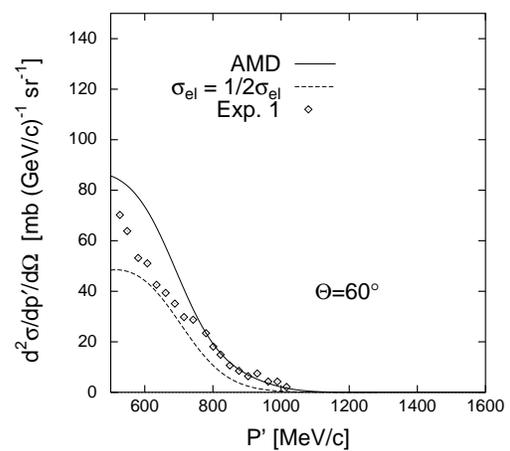
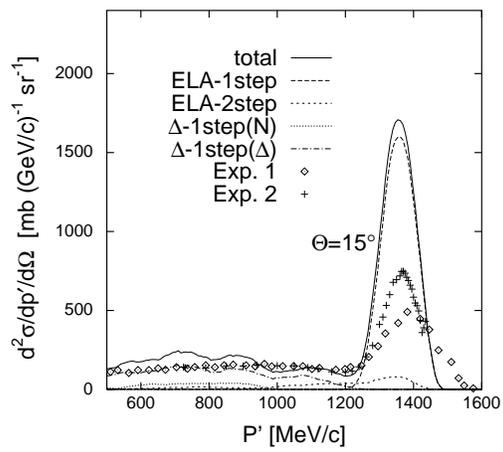
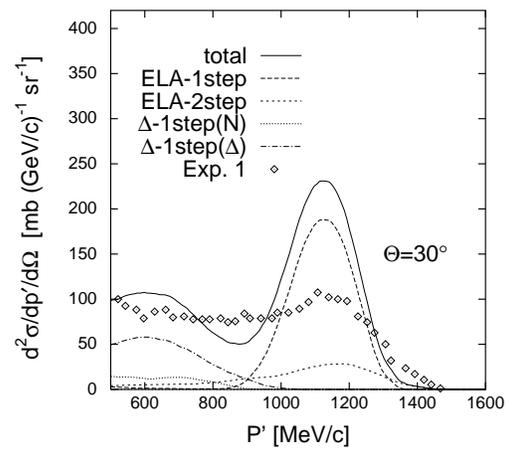
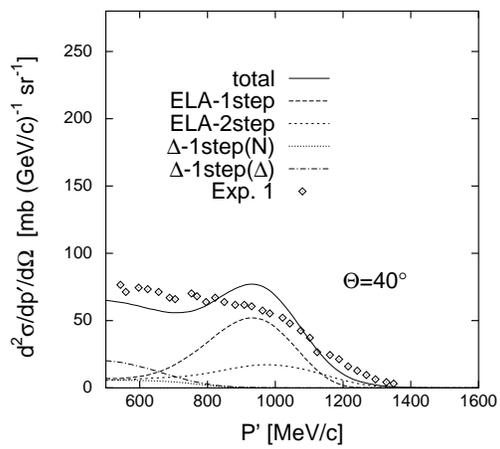
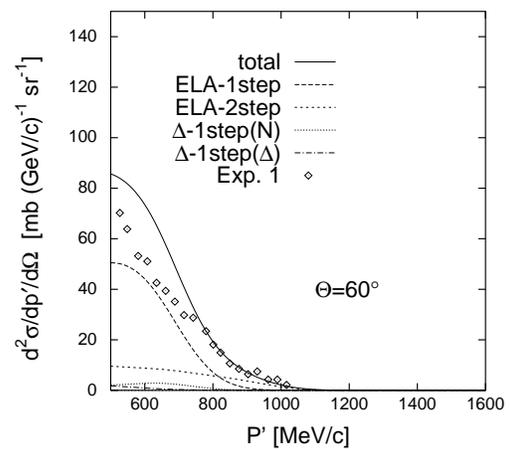
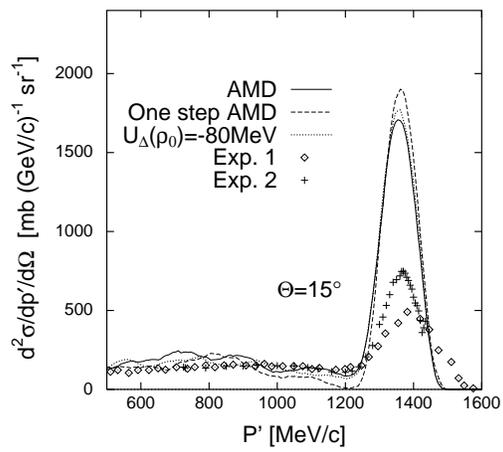
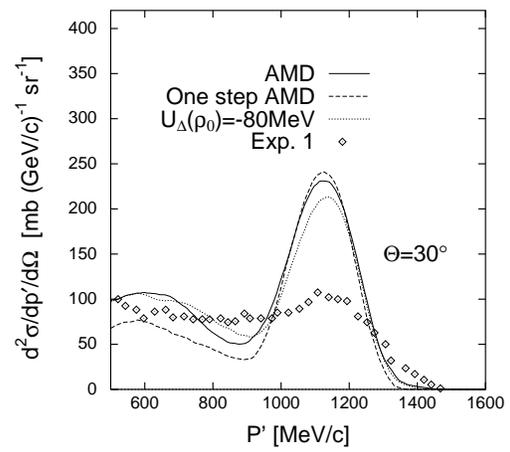
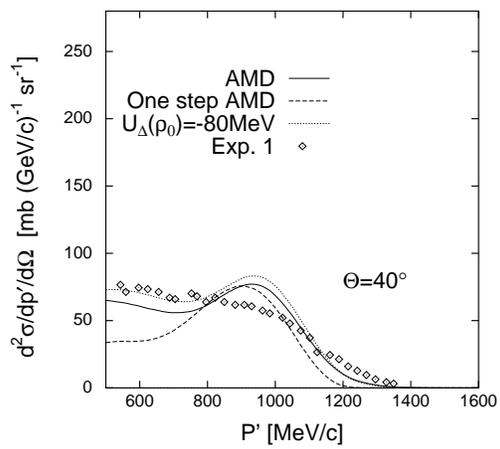
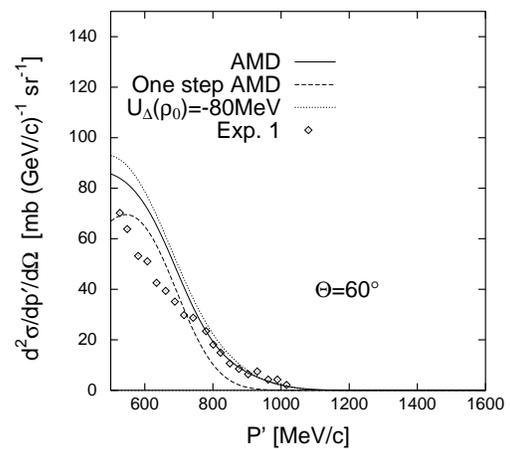
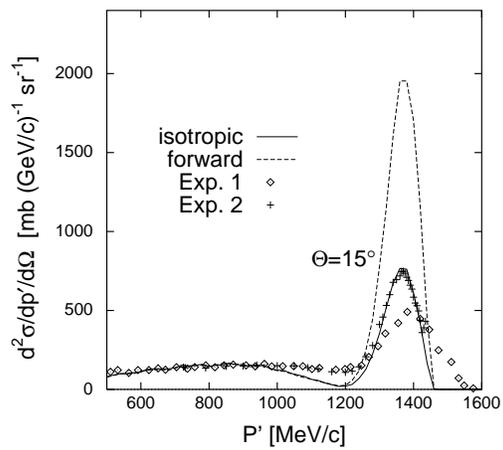
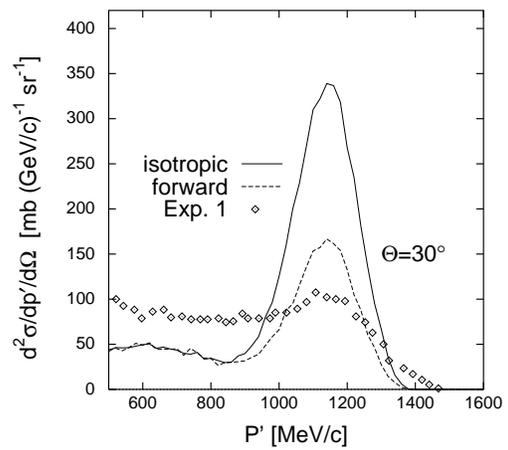
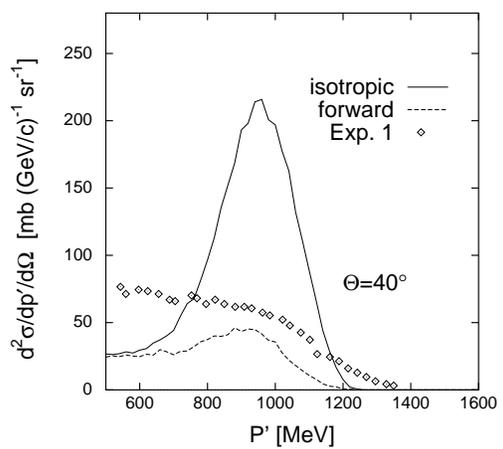
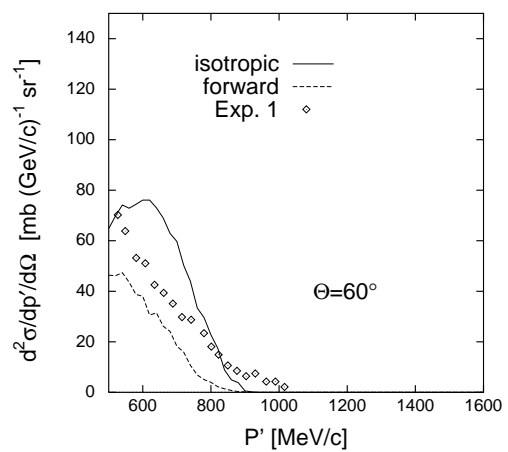
\end{document}